\documentclass[aps,preprintnumbers,superscriptaddress]{revtex4-1}
\usepackage{amsmath, mathrsfs, amssymb,amsfonts,amsthm,graphicx, epsf, dcolumn, yfonts}
\usepackage[normalem]{ulem}
\usepackage[hyperfootnotes=true]{hyperref}
\usepackage{color}
\usepackage{amssymb}
\usepackage{slashed}
\usepackage{setspace}
\usepackage{cancel}
\usepackage{wasysym}
\usepackage{tikz-feynman}
\usepackage{float}
\usepackage[utf8]{inputenc}
\usepackage{todonotes}
\usepackage{dsfont}
\usepackage{amsmath}
\usepackage{tikz}
\usepackage{tikz-cd}

\usepackage{subcaption}
\usepackage{array}
\usepackage{tabularray}
\usepackage{tabularx}
\usepackage{ragged2e}

        \hypersetup{
           breaklinks=true,   
           colorlinks=true,   
           pdfusetitle=true,  
        }
        
\pdfoutput=1
\definecolor{bottle_green}{RGB}{0,106,78}
\definecolor{celadon_green}{RGB}{47,132,124}
\definecolor{emerald}{RGB}{80,220,100}
\definecolor{jade}{RGB}{0,168,107}
\hypersetup{colorlinks,
linkcolor=bottle_green,
citecolor=celadon_green,
urlcolor=emerald,
anchorcolor=jade}

\newcommand{\q}{q}

\def\<{\langle}
\def\>{\rangle}

\newcommand{\be}{\begin{eqnarray} \begin{aligned}}
\newcommand{\ee}{\end{aligned} \end{eqnarray} }
\newcommand{\benn}{\begin{eqnarray*} \begin{aligned}}
\newcommand{\eenn}{\end{aligned} \end{eqnarray*} }

\newcommand{\ben}{\begin{eqnarray} \begin{aligned}}
\newcommand{\een}{\end{aligned} \end{eqnarray} }

\newcommand{\bc}{\begin{center}}
\newcommand{\ec}{\end{center}}

\newcommand{\beq}{\begin{eqnarray} \begin{aligned}}
\newcommand{\eeq}{\end{aligned} \end{eqnarray} }
\newcommand{\bea}{\begin{array}}
\newcommand{\eea}{\end{array}}

\newcommand{\bee}{\begin{enumerate}}
\newcommand{\eee}{\end{enumerate}}
\newcommand{\bei}{\begin{itemize}}
\newcommand{\eei}{\end{itemize}}

\usepackage{amsfonts}

\def\01{\{0,1\}}

\newcommand{\cE}{\mathcal{E}}

\newcommand{\str}{{\vec{x}}}

\def\<{\langle}
\def\>{\rangle}

\def\s{\,\,\,\,}

\newtheorem*{rep@theorem}{\rep@title}
\newcommand{\newreptheorem}[2]{%
\newenvironment{rep#1}[1]{%
 \def\rep@title{#2 \ref{##1} (restatement)}%
 \begin{rep@theorem}}%
 {\end{rep@theorem}}}
\makeatother

\newreptheorem{thm}{Theorem}
\newreptheorem{lem}{Lemma}

\definecolor{ma}{RGB}{246,76,246}
\definecolor{re}{RGB}{210,42,42}

\def\T{\bf T}

\def\z{{ z}}

\def\g{{g}}

\def\T00{{\bf T_{NN}}}

\def\0mom{{\bar{\Gamma}^{\alpha\beta}(\z)}}
\def\1mom{{\Gamma^{\alpha\beta}_1(\z)}}
\def\2mom{{\Gamma^{\alpha\beta}_2(\z)}}

\begin{document}

\title{A quantum oscillator interacting with a classical oscillator}

\author{Muhammad Sajjad}
\affiliation{Department of Physics, Middle East Technical University, 06800, Ankara, Turkiye}

\author{Andrea Russo}
\affiliation{Department of Physics and Astronomy, University College London, Gower Street, London WC1E 6BT, United Kingdom}

\author{Maite Arcos}
\affiliation{Department of Physics and Astronomy, University College London, Gower Street, London WC1E 6BT, United Kingdom}

\author{Andrzej Grudka}
\affiliation{Institute of Spintronics and Quantum Information, Faculty of Physics, Adam Mickiewicz University, 61-614 Poznań, Poland}

\author{Jonathan Oppenheim}
\affiliation{Department of Physics and Astronomy, University College London, Gower Street, London WC1E 6BT, United Kingdom}

%
%

\begin{abstract}
We study a quantum oscillator interacting with, and back-reacting on, a classical oscillator. This can be done consistently provided the quantum system decoheres, while the backreaction has a stochastic component which causes the classical system to undergo diffusion. Nonetheless, the state of the quantum oscillator can remain pure conditioned on the trajectory of the classical oscillator.
We solve the system using the classical-quantum path integral formulation, and compute the free and bridge correlators of the theory, which we use to find the heating rate. In addition, we explicitly compute the conditioned, and unconditional purity of the system. This serves as a toy model for a number of other systems in which one subsystem can be treated as effectively classical, such as a scalar quantum field interacting with another field undergoing decoherence, or a system emitting radiation, one of which is treated classically.
\end{abstract}

\maketitle

\section{Introduction}
\label{sec: intro}

Classical-quantum systems are ubiquitous in modern-day physics. When the classical system influences the quantum system, we can easily account for this through external parameters in the Hamiltonian of the quantum system. Common examples include an atom under the influence of a classical electromagnetic field, and attosecond physics, where nuclear motion is often treated classically whilst the electron is treated quantum mechanically~\cite{attosecond}. Quantum field theory in curved space is another such example. The back-reaction of quantum systems on classical ones is harder to account for, at least if one wants to be consistent.
Such situations often arise in quantum chemistry, hybrid quantum computers and quantum technologies, light-initiated energy transfer in biology~\cite{QuantumBiology}, quantum thermodynamics where the heat bath is treated classically, and quantum control with feedback. They are also relevant to understanding gravity, as we currently lack a complete theory of quantum gravity. Here, we would like to understand phenomena such as black-hole evaporation, where the emitted radiation is quantum in nature, yet the spacetime surrounding the black hole can only be described classically. Additionally, we would like to understand how vacuum fluctuations contribute to the expansion factor during the big bang and to structure formation, since the former are quantum while the latter we only know how to describe classically.

In some cases, one may be aiming for an approximation scheme in which negative probabilities can be tolerated~\cite{tully1990molecular,tully1998mixed,kapral1999mixed}. But if one wants to respect positivity and normalisation, then consistency requires that the dynamics be a completely positive and trace-preserving (CPTP) map. If the underlying system is genuinely quantum, then both approaches have different advantages; here we address the CPTP case.
Early examples of consistent quantum-classical master equations were given in the 1990s~\cite{blanchard1995event,diosi1995quantum}, where the collisionless Boltzmann equation~\cite{alicki2003completely} and toy models for gravity~\cite{diosi2011gravity,poulinKITP} were studied. A weak-measurement and feedback approach has been used to study models of a classical Newtonian potential sourced by measurements or spontaneous collapse models~\cite{kafri2014classical,tilloy2016sourcing, tilloy2017principle,diosi1998coupling}. Recently, the most general form of such dynamics was derived~\cite{oppenheim2018post,UCLPawula}, and used to construct general trajectories~\cite{UCL2022semi} and a path integral~\cite{oppenheim2023path} for classical-quantum dynamics. This has been applied to construct a general relativistic theory in which the spacetime metric is treated classically, while matter is described by quantum fields~\cite{oppenheim2018post,oppenheim2023covariant,layton2023weak,oppenheim2024diffeomorphism}.


In~\cite{layton2023classical}, it was shown that consistent classical-quantum dynamics can be obtained as the classical limit of a doubly quantum system, in which the classical limit is taken for only one of the two subsystems. This highlights the potential of the approach as an effective theory of classical-quantum dynamics in settings such as chemistry, molecular biology, and thermodynamical and gravitational systems.
%
%
However, we would like to study this limit in the simplest possible systems. The case of a constant force was first studied in~\cite{diosi1995quantum}, and later in~\cite{UCLqubit}. Here, we wish to go beyond the constant-force case and study the continuous dynamics of interacting harmonic oscillators.
These serve as a standard toy model for anything from interacting quantum fields to stars emitting radiation.
The master equation for a classical oscillator interacting with a quantum one was given in~\cite{UCLPawula}, but not studied. In this work, we apply the path integral approach~\cite{oppenheim2023path} to the interaction of a quantum harmonic oscillator with a classical one. Since the path integral is quadratic, we are able to solve it completely.
A discussion of the classical-quantum oscillator using other classical-quantum approaches can be found in~\cite{terno2024classicalquantum}.

Details regarding the path integral approach to classical-quantum dynamics can be found in~\cite{oppenheim2023path,oppenheim2023covariant}. The quantity of interest is
\begin{align}
    \varrho(z_f,\phi^+_f,\phi^-_f,t_f) &= \int \mathcal{D}z\, \mathcal{D} \phi^+\mathcal{D}\phi^- \;\mathcal{N}e^{\mathcal{I}_{CQ}(z,\phi^+,\phi^-,t_i,t_f)}\, \varrho(z_i,\phi^+_i,\phi^-_i,t_i),
    \label{eq: CQPI}
\end{align}
where $z$ are the classical degrees of freedom and $\phi^\pm$ are the bra and ket degrees of freedom; we double the quantum degrees of freedom as in the Feynman--Vernon path integral, since we are computing not an amplitude but a density matrix. The subscript $f$ indicates the values of the degrees of freedom at time $t_f$, and the path integral computes the hybrid classical quantum (CQ) state $\varrho(z,\phi^+,\phi^-,t)$ as it evolves from the initial to the final time. $\mathcal{N}$ is the path integral normalisation, and the probability weighting is given by the classical-quantum action $\mathcal{I}_{CQ}$, defined as
\begin{align}
      \mathcal{I}_{CQ}(z,\phi^+,\phi^-,t_i,t_f) &=\int_{t_i}^{t_f} dt \bigg[ i\big(\mathcal{L}_Q(\phi^+)-\mathcal{L}_Q(\phi^-)\big) -\frac{1}{2}\frac{ \delta  \Delta W_{CQ}}{ \delta z_{i}} D_{0,ij} \frac{\delta \Delta W_{CQ}}{ \delta z_{j}} \notag\\
      &- \frac{1}{2} \frac{\delta \bar{W}_{CQ} }{ \delta z_i} D_{2, ij}^{-1} \frac{\delta \bar{W}_{CQ} }{ \delta z_j}\bigg],
      \label{eq: CQAction}
\end{align}
where we have set $\hbar=1$. Here $\mathcal{L}_Q(\phi^+)$ and $\mathcal{L}_Q(\phi^-)$ are the usual Lagrangian of the quantum system evaluated on the bra and ket quantum degrees of freedom. $W_{CQ}$ is called the \textit{proto-action} (with $\bar{W}_{CQ}\equiv\frac{1}{2}(W_{CQ}(\phi^{+})+W_{CQ}(\phi^{-})), \Delta W_{CQ}\equiv W_{CQ}(\phi^{+})-W_{CQ}(\phi^{-})$) and encodes the Lagrangian of the classical degrees of freedom plus any interaction term with the quantum ones. The first two terms of the path integral represent the unitary evolution and the decoherence of the quantum system, which is regulated by the \textit{decoherence coefficient} $D_{0,ij}$, while the last term encodes the diffusion of the classical degrees of freedom, regulated by the \textit{diffusion coefficient} $D_{2,ij}$. All of this will become clearer in the next section, where we apply the path integral to a classical-quantum oscillator.  we set $\hbar=1$.

In Section~\ref{sec: SHO_PI}, we start by constructing the path integral~\eqref{eq: PI_SHO} and the action~\eqref{eq: PI_SHO_ACTION} for the CQ harmonic oscillator; this makes evident the roles of the different terms in the path integral. Afterwards, in Section~\ref{sec: greens}, we  compute the matrix of Green's functions explicitly \eqref{eq: M_inverse} and justify the boundary conditions used. This computation is exact since the path integral is Gaussian. Then, in Section~\ref{sec: Bridge}, we compute the semi-classical action \eqref{eq: S_onshell_3}, and solve the equations of motion for the classical and quantum variables. This allows us to turn our attention to Brownian bridge case, i.e., where both the initial and final conditions of the classical system are given and used to find weak expectation values \cite{aharonov1988result}. In \ref{sec: Heating} we are thus able to compute the expectation value of energy for both the free and conditioned cases, and show that the this system does not have an equilibrium energy. Finally, in \ref{sec: measurement}, we recast the hybrid harmonic oscillator problem as a measurement and feedback one, and compute the purity conditioned on a classical state, as well as the unconditioned purity.

\section{Path integral for the hybrid harmonic oscillators}
\label{sec: SHO_PI}

In this section, we construct the path integral for the CQ harmonic oscillators following the procedure delineated in~\cite{oppenheim2023path,oppenheim2023covariant}. The quantum degree of freedom is a harmonic oscillator $Q$ with frequency $\omega_Q$, such that the bra and ket branches are represented by $(\phi^+,\phi^-)=(Q^+,Q^-)$ (which we sometimes write with the plus and minus as subscripts for clarity), while the classical degree of freedom is taken to be that of a classical stochastic harmonic oscillator in configuration space, with position $z=q$ and frequency $\omega_c$. The Lagrangians of the two systems are given by
\begin{align}
    \mathcal{L}_c(q,\dot{q})&=\frac{1}{2}\dot{q}^2-\frac{1}{2}\omega_c^2 q^2, \label{eq: L_c}\\
    \mathcal{L}_Q(Q,\dot{Q})&=\frac{1}{2}\dot{Q}^2-\frac{1}{2}\omega_Q^2 Q^2-\mathcal{V}_i(q,Q), \label{eq: L_Q}
\end{align}
where we have completed the quantum Lagrangian with an interaction potential. In general, one could explore the consequences of non-linear interactions between the quantum and classical systems but, for the scope of this work, we restrict the potential to a linear one,
\begin{align}
    \mathcal{V}_i&=\alpha q Q. \label{eq: V_int}
\end{align}
 This coupling~\eqref{eq: V_int} is the natural choice for several reasons. A coupling through the distance between the oscillators reduces to it:
  $\tfrac{k}{2}(q-Q)^2=\tfrac{k}{2}q^2+\tfrac{k}{2}Q^2-k\,qQ$, where the first two terms only shift $\omega_c^2$ and $\omega_Q^2$ and the last is our
  coupling with $\alpha=-k$.  More generally, a bilinear term is the lowest order at which a smooth interaction $\mathcal V(q,Q)$, expanded about
  equilibrium, couples the two oscillators. Further, it is the only coupling for which the CQ action remains quadratic,
  so that the path integral is Gaussian and every result below is exact, at any coupling strength. Thus, in the Calderia-Leggett model, for example, the quantum system couples linearly to the bath. A more general potential changes the physics in two
  ways. First, the back-reaction force on the classical oscillator is $-\partial_q\mathcal V$, and complete positivity then requires decoherence in the
  eigenbasis of $\partial_q\mathcal V$ rather than of $Q$: for $\mathcal V=q\,f(Q)$ the quantum oscillator decoheres in the eigenbasis of $f(Q)$, so that
  for $f(Q)=Q^2$ superpositions of $\pm Q$ are not decohered at all. Second, if $\partial_q\mathcal V$ depends on $q$, the decoherence and diffusion become
  state-dependent. A non-linear coupling also makes the action non-Gaussian. For weak coupling, such interactions can be treated perturbatively about the
  Gaussian theory, whose exact propagators we compute here.

We can form the proto-action as indicated in~\cite{oppenheim2023covariant} by completing the classical Lagrangian with the interaction potential:
\begin{align}
    W_{CQ}(q,Q)&=\mathcal{L}_c(q,\dot{q})-\mathcal{V}_i(q,Q) \notag\\
    &=\frac{1}{2}\dot{q}^2-\frac{1}{2}\omega_c^2 q^2-\alpha qQ. \label{eq: protoaction}
\end{align}
The oscillators have unit mass. Here we chose a spontaneous localisation model such that $D_{0,ij}=D_0\delta_{ij}$ and $D^{-1}_{2,ij}=D_2^{-1}\delta_{ij}$ with $D_0$ and $D_2$ positive definite.
The configuration-space CQ path integral of Equation~\eqref{eq: CQPI} then takes the form
\begin{align}
    \varrho(q_f,Q^+_f,Q^-_f,t_f) &= \int \mathcal{D}q\, \mathcal{D} Q^+\mathcal{D}Q^- \;\mathcal{N}e^{\mathcal{I}_{CQ}[q,Q^+,Q^-,t_i,t_f]}\, \varrho(q_i,Q^+_i,Q^-_i,t_i),
    \label{eq: PI_SHO}
\end{align}
with action given by
\begin{align}
      \mathcal{I}_{CQ}(q,Q^+,Q^-,t_i,t_f) &=\int_{t_i}^{t_f} dt \bigg[ i\left(\frac{1}{2}\dot{Q}^2_+ -\frac{1}{2}\omega_Q^2 Q^2_+-\alpha q Q_+\right)-i\left(\frac{1}{2}\dot{Q}^2_- -\frac{1}{2}\omega_Q^2 Q^2_--\alpha q Q_-\right) \notag\\
      &- \frac{D_0\,\alpha^2}{2}(Q_+-Q_-)^2-\frac{D_2^{-1}}{2}\left(\ddot{q}+\omega^2_c q+\frac{\alpha}{2}(Q_+ + Q_-)\right)^2\bigg].
      \label{eq: PI_SHO_ACTION}
\end{align}

This action is more illuminating when rewritten in the Schwinger--Keldysh basis,
\begin{align}
Q_c&=\frac{1}{2}(Q^{+}+Q^{-}), \qquad Q_q=Q^{+}-Q^{-}, \qquad Q^{\pm}=Q_c\pm\frac{1}{2}Q_q, \label{eq: Keldysh_basis}
\end{align}
in which it reads
\begin{align}
 \mathcal{I}_{CQ}(q,Q_c,Q_q,t_i,t_f)&= \int_{t_i}^{t_f} dt\Big[ i\Big(\dot Q_c\dot Q_q-\omega_Q^2Q_{c}Q_q-\alpha qQ_q\Big)
  -\frac{D_0\alpha^2}{2}Q_q^2
  -\frac{D_2^{-1}}{2}\Big(\ddot q+\omega_c^2q+\alpha Q_c\Big)^2\Big]. \label{eq: PI_SHO_ACTION_Keldysh}
\end{align}
In this form, the role of each term is easy to see:
\begin{align}
\operatorname{Re}\mathcal{I}_{CQ}[Q_c,Q_q,q]&=-\frac{\alpha^2D_0}{2}\int_0^t\!ds\,Q_q^2
  -\frac{1}{2D_2}\int_0^t\!ds\,\big(\ddot q+\omega_c^2q+\alpha Q_c\big)^2 . \label{eq: ReAction}
\end{align}
$Q_q$ enters the real part only through the first term, which causes the decay of the off-diagonal terms, i.e., the primary decoherence.
The second term is the classical system's Onsager--Machlup weight, and it suppresses histories of $Q_c$
that move away from the classical, noiseless trajectory. It therefore acts on the diagonal of the quantum system: it correlates
$Q_c$ with $q$, but it does not by itself damp $Q_q$. $Q_q$ couples to the meter only through the
phase $i\alpha\int Q_q\,q$. Integrating over $q$ with the Gaussian weight of the second term turns
this phase into the real, negative-definite functional $-\frac{\alpha^2}{2}\iint Q_q\,\mathcal N\,Q_q$, with $\mathcal{N}$ given in Eq.~\eqref{eq: noise_kernel}.
This decoherence attributed to $q$ is the secondary decoherence, visible only
after $q$ is traced out. Note that the real part is negative definite and therefore does not lead to unbounded probabilities.

\emph{The diffusion-decoherence trade-off}: For complete positivity, the cross terms $Q^{+}Q^{-}$ must have an overall positive coefficient, which one can verify by performing a short-time expansion. This gives $\alpha^2(D_0-\frac{1}{4}D_2^{-1})\geq 0 \Rightarrow 4D_0 \geq D_2^{-1}$. Thus, the amount of stochasticity in the classical oscillator sets how fast the quantum oscillator must decohere. The less noisy the classical oscillator, the faster the quantum oscillator decoheres. In the limit $D_2\to0$, where the classical oscillator is completely deterministic, the decoherence rate diverges and the quantum oscillator decoheres instantaneously. Additionally, the classical system boosts anti-correlations between $Q^{+}$ and $Q^{-}$, but the decoherence term counteracts these effects. At saturation, $D_0=\frac{1}{4}D_2^{-1}$, the cross terms vanish, and a pure state, conditioned on a classical trajectory, remains pure (see also Section~\ref{sec: measurement}).

\emph{Non-Markovianity and the decoherence rate}: The action as given is clearly Markovian. However, integrating out the classical system gives the following effective action:
\begin{align}
  \operatorname{Re}\mathcal{I}_{\rm eff}[Q_q]
  &=-\frac{\alpha^2D_0}{2}\int_0^t\!ds\,Q_q(s)^2
  -\frac{\alpha^2}{2}\int_0^t\!\!\int_0^t\!ds\,ds'\,Q_q(s)\,\mathcal N(s,s')\,Q_q(s'), \label{eq: ReAction_eff}
\end{align}
with
\begin{align}
  \mathcal N(s,s')&=D_2\int_0^{\min(s,s')}\!du\,\frac{\sin\omega_c(s-u)}{\omega_c}\,\frac{\sin\omega_c(s'-u)}{\omega_c} \label{eq: noise_kernel}
\end{align}
where $\mathcal{N}$ is the Green's function of the operator $D_2^{-1}(\partial_t^2+\omega_c^2)^2$ with clamped-free boundary conditions.
Equivalently, the second term is $-\frac{\alpha^2D_2}{2}\int_0^t du\,\Theta(u)^2$, where
\begin{align}
  \Theta(u)&=\int_u^t ds\,\frac{\sin\omega_c(s-u)}{\omega_c}\,Q_q(s) \label{eq: Theta}
\end{align}
is the advanced smearing of $Q_q$. Both terms are negative definite in $Q_q$, and together they set the decoherence rate of the system.

  \begin{figure}[t]
    \centering
    \includegraphics[width=\linewidth]{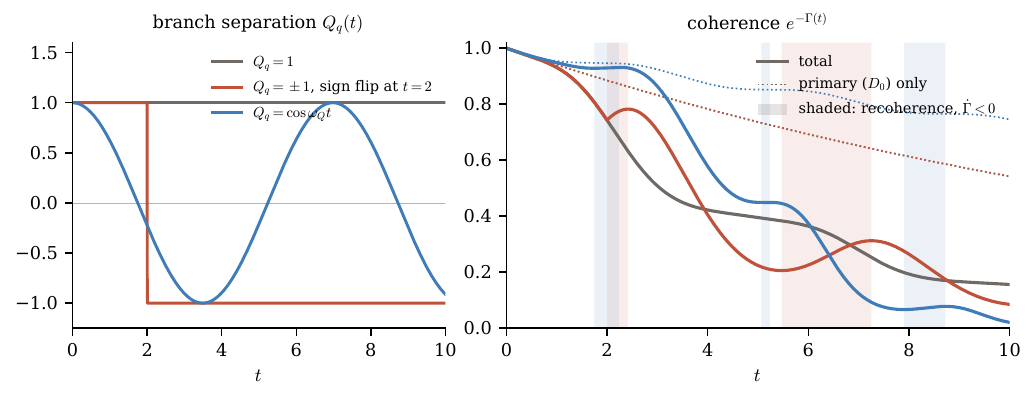}
    \caption{Decoherence and recoherence of the quantum oscillator, obtained from the effective action~\eqref{eq: ReAction_eff} by integrating the
  rate~\eqref{eq: Gamma_rate_factorised}, for $\alpha=0.7$, $D_0=0.25$, $D_2=1$, $\omega_c=1.3$ and $\omega_Q=0.9$. Left: three histories of the separation
  $Q_q=Q^+-Q^-$ of the two branches: constant (grey), changing sign at $t=2$ (red), and oscillating at the frequency of the quantum oscillator,
  $Q_q=\cos\omega_Qt$ (blue). Right: the coherence factor $e^{-\Gamma(t)}$ (solid), and the contribution of the primary decoherence alone,
  $\exp\!\big(-\tfrac{\alpha^2D_0}{2}\int_0^tQ_q^2\big)$ (dotted; it is the same for the first two histories). For constant $Q_q$ the coherence decays
  monotonically. When $Q_q$ changes sign, the memory of the classical oscillator partially undoes the decoherence it has caused, and $e^{-\Gamma}$ increases
  in the shaded intervals, where $Q_q(t)$ lies strictly between $0$ and $-2M(t)/D_0$. For the sign change at $t=2$ the coherence recovers by a factor
  $e^{0.42}\simeq1.5$. The primary contribution never increases.}
    \label{fig: decoherence}
  \end{figure}

\emph{The decoherence rate}:
  Defining the decoherence exponent $\Gamma(t)\equiv-\operatorname{Re}\mathcal{I}_{\rm eff}$ and using the symmetry $\mathcal N(s,s')=\mathcal N(s',s)$, the
  rate is
  \begin{align}
    \frac{d\Gamma}{dt}&=\alpha^2\,Q_q(t)\Big[\frac{D_0}{2}\,Q_q(t)+M(t)\Big], \label{eq: Gamma_rate_factorised}
  \end{align}
where we define the memory, $M(t)\equiv\int_0^t\!ds'\,\mathcal N(t,s')\,Q_q(s')$.
  The first term is non-negative: the primary decoherence never decreases. The second term, however, has no definite sign. Complete positivity guarantees
  $\Gamma(t)\geq0$, since $\mathcal N$ is a positive kernel, but not that $\Gamma$ increases. The quantum oscillator therefore recoheres at time $t$ if and only if $Q_q(t)$ lies strictly between $0$ and $-2M(t)/D_0$. For a constant $Q_q$, the second term equals
  $\alpha^2D_2\,Q_q^2\,(1-\cos\omega_c t)^2/(2\omega_c^4)\geq0$ and the decoherence is monotone.

\subsection{Green's functions}
\label{sec: greens}

One can now invert the action to find the matrix of Green's functions. Writing the action as
\begin{align}
    \mathcal{I}_{CQ}&=-\frac{1}{2}\int^{t_f}_{t_i}dt\, \vec{\phi}^{T}\mathcal{M}\vec{\phi}, \label{eq: action_quadratic}
\end{align}
with $\vec{\phi}^{T}=(Q_{c},Q_q,q)$, one has symbolically
\begin{align}
  \mathcal{M}&=  \begin{bmatrix}
    \frac{\alpha^2}{D_2}&iL_Q^{A}&\frac{\alpha}{D_2}L_q^R \\iL_Q^R&\alpha^2D_0&i\alpha\\ \frac{\alpha}{D_2}L_q^A&i\alpha&\frac{1}{D_2}L_q^A \circ L_q^R
    \end{bmatrix}, \label{eq: M_matrix}
\end{align}
where $L_{Q/q}^{R/A}=(\partial_t^2+\omega^2_{Q/c})^{R/A}$, i.e., the operator with retarded or advanced boundary conditions. Inverting, one obtains
\begin{align}
    \mathcal{M}^{-1}&= \frac{1}{\Delta^{R}\Delta^{A}}
    \begin{bmatrix}
      \alpha^{2}\Big(D_0\,L_q^{A}\!\circ\!L_q^{R}+D_2\Big)
        & -i\,L_q^{R}\,\Delta^{A}
        & -\alpha\Big(D_2\,L_Q^{A}+\alpha^{2}D_0\,L_q^{R}\Big)\\[2mm]
      -i\,L_q^{A}\,\Delta^{R}
        & 0
        & i\alpha\,\Delta^{R}\\[2mm]
      -\alpha\Big(D_2\,L_Q^{R}+\alpha^{2}D_0\,L_q^{A}\Big)
        & i\alpha\,\Delta^{A}
        & D_2\,L_Q^{A}\!\circ\!L_Q^{R}+\alpha^{4}D_0
    \end{bmatrix}, \label{eq: M_inverse}
\end{align}
where $\Delta^A\equiv L_Q^AL_q^A-\alpha^2$, and similarly for $\Delta^R$.

Reading off the two-point functions, we have
\begin{align}
  \langle Q_{c}Q_{c}\rangle&=\frac{\alpha^{2}\big(D_0 L_q^{A}\circ L_q^{R}+D_2\big)}{\Delta^{R}\Delta^{A}}, &
  \langle qq\rangle&=\frac{D_2L_Q^{A}\circ L_Q^{R}+\alpha^{4}D_0}{\Delta^{R}\Delta^{A}}, &
  \langle Q_q Q_q\rangle&=0, \label{eq: 2pf_op_1}\\
  \langle qQ_c\rangle&=\frac{-\alpha\big(D_2L_Q^{R}+\alpha^{2}D_0L_q^{A}\big)}{\Delta^{R}\Delta^{A}}, &
  \langle qQ_q\rangle&=\frac{i\alpha}{\Delta^{R}}, &
  \langle Q_cQ_q\rangle&=-i\frac{L_q^{R}}{\Delta^{R}}, \label{eq: 2pf_op_2}
\end{align}
with $R\leftrightarrow A$ for the transposes.

\emph{Deriving the prescription}: Let us now briefly describe how the prescriptions in $\mathcal{M}$ were obtained. Conceptually, one divides the fields into a semi-classical part and a fluctuation part, where the semi-classical part is defined as the one that extremises the action. The global extremum is given by the solution of the following equations:
\begin{align}
\ddot{q}+\omega_c^2q+\alpha Q_c&=0, \label{eq: sc_eom_1}\\
Q_q&=0, \label{eq: sc_eom_2}\\
\ddot{Q}_c+\omega_Q^2Q_c+\alpha q&=0. \label{eq: sc_eom_3}
\end{align}
With the boundary conditions $Q_q(t_f)=\dot{Q}_q(t_f)=0$, one notes that this gives $\mathcal{I}_{CQ,{\rm cl}}=0$. The matrix of Green's functions is then the inverse of the matrix of fluctuations. The boundary conditions are set by causality: the quantum sector has the usual advanced/retarded structure, while for the classical sector the $\mathcal{M}_{qq}$ prescription is set by clamped-free boundary conditions: the field and its derivative are fixed at $t_i$ but are free at $t_f$, where, instead, $L_q q$ and $\partial_t(L_q q)$ are fixed. Similarly, the classical-quantum entries are fixed by causality: in the real part of the action, the classical sector back-reacts on the quantum one, hence the retarded prescription for $\mathcal{M}_{Q_{c}q}$.

As a check, one can instead turn to a Martin--Siggia--Rose (MSR) formulation \cite{janssen1976lagrangean} \cite{martin1973statistical} \cite{dedominicis1976techniques} for the classical field, using a Hubbard-Stratonovich transformation:
\begin{align}
    \mathcal{I}^{\rm MSR}_{CQ}&=\int_{t_i}^{t_f} dt\Big[ i\Big(\dot Q_c\dot Q_q-\omega_Q^2Q_{c}Q_q-\alpha qQ_q\Big)
  -\frac{D_0\alpha^2}{2}Q_q^2
  +i\tilde{q}\Big(\ddot q+\omega_c^2q+\alpha Q_c\Big)-\frac{D_2}{2}\tilde{q}^2\Big]. \label{eq: MSR_action}
\end{align}
This formulation lends itself more naturally to causality conditions: the initial conditions for the field propagate forward in time, while the final conditions for the response field propagate backwards. Integrating out the response field, one obtains the same prescription as above.

\subsection{Time representation}
\label{sec: time_rep}
To write down the explicit form of the two-point functions, we first compute the Green's function $g^{R/A}(t,s)$ of $\Delta^{R/A}\equiv L_Q^{R/A}L_q^{R/A}-\alpha^2$:
\begin{align}
  g^{R}(t,s)&=\frac{\theta(t-s)}{\omega_+^2-\omega_-^2}\left[\frac{\sin\omega_-(t-s)}{\omega_-}-\frac{\sin\omega_+(t-s)}{\omega_+}\right], \label{eq: gR}
\end{align}
with $t$ and $s$ interchanged in the advanced case. Here $\omega_\pm^2$ are obtained by factorising $\Delta$,
\begin{align}
  \Delta&=\big(\partial_t^2+\omega_Q^2\big)\big(\partial_t^2+\omega_c^2\big)-\alpha^2=\big(\partial_t^2+\omega_+^2\big)\big(\partial_t^2+\omega_-^2\big), \label{eq: Delta_factor}\\
  \omega_\pm^2&=\frac{\omega_Q^2+\omega_c^2}{2}\pm\sqrt{\Big(\frac{\omega_Q^2-\omega_c^2}{2}\Big)^2+\alpha^2}. \label{eq: omega_pm}
\end{align}
For $\tau=t-s>0$,
\begin{align}
  L_{q/Q}\,g^{R}&=\frac{1}{\omega_+^2-\omega_-^2}\left[\big(\omega_+^2-\omega_{c/Q}^2\big)\frac{\sin\omega_+\tau}{\omega_+}+\big(\omega_{c/Q}^2-\omega_-^2\big)\frac{\sin
  \omega_-\tau}{\omega_-}\right]. \label{eq: Lg}
\end{align}
Using these, we have
\begin{align}
\langle qQ_q\rangle&={i\alpha}\,g^R(t,s), \qquad
  \langle Q_cQ_q\rangle=-i\,{L_q^{R}}g^R(t,s), \label{eq: 2pf_response}
\end{align}
both of which are bounded, time-translation-invariant functions in the subcritical case, $\alpha^2<\omega_c^2\omega_Q^2$.\footnote{In the critical and supercritical cases all two-point functions are unbounded.}

The other two-point functions, $\langle qq\rangle$, $\langle qQ_c\rangle$ and $\langle Q_c Q_c\rangle$, however, grow secularly with time. To compute them, we use $(\Delta^{R}\Delta^A)^{-1}=(|\Delta|^2)^{-1}$, which is given by
\begin{align}
  \frac{1}{|\Delta|^{2}}&=\big(g^R\circ g^{R,\mathsf T}\big)(t,t')
  =\int_{0}^{\min(t,t')}\!ds\;g(t-s)\,g(t'-s), \label{eq: Delta_inv_sq}
\end{align}
where the lower limit is set by the initial time, $t_i=0$, and the upper limit by the Heaviside functions $\theta(t-s)\theta(t'-s)$, and we have used $g^A=g^{R,\mathsf T}$.

To evaluate this expression, we use
\begin{align}
  J_{ab}(t,t')&\equiv\int_0^{\min(t,t')}\!\!\!ds\;
  \frac{\sin\omega_a(t-s)}{\omega_a}\;\frac{\sin\omega_b(t'-s)}{\omega_b},
  \qquad a,b\in\{+,-\}, \label{eq: J_def}
\end{align}
which evaluates to
\begin{align}
  J_{aa}(t,t')&=\frac{\min(t,t')}{2\omega_a^{2}}\cos\omega_a(t-t')
  +\frac{\sin\omega_a|t-t'|-\sin\omega_a(t+t')}{4\omega_a^{3}}, \label{eq: J_aa}\\
  J_{ab}(t,t')&=\frac{1}{2\omega_a\omega_b}\Bigg[
  \frac{\sin\big(\omega_at-\omega_bt'\big)
  -\sin\big(\omega_a(t-m)-\omega_b(t'-m)\big)}{\omega_a-\omega_b} \notag\\
  &-\frac{\sin\big(\omega_at+\omega_bt'\big)
  -\sin\big(\omega_a(t-m)+\omega_b(t'-m)\big)}{\omega_a+\omega_b}
  \Bigg], \qquad m=\min(t,t'),\quad a\neq b. \label{eq: J_ab}
\end{align}
It is $J_{aa}$ that is responsible for the secular growth.

The two-point functions are then given by
\begin{align}
  \langle q(t)q(t')\rangle\equiv C_{qq}(t,t')&=\int_0^{\min(t,t')}\!\!\!ds\;\Big[\alpha^4D_0\;g(t-s)\,g(t'-s)\notag\\
  &+D_2\,\big(L_Qg\big)(t-s)\,\big(L_Qg\big)(t'-s)\Big]\notag\\
  &=\frac{1}{\big(\omega_+^2-\omega_-^2\big)^2}
  \Big\{\alpha^4D_0\Big[J_{++}-J_{+-}-J_{-+}+J_{--}\Big]\notag\\
  &+D_2\Big[\big(\omega_c^2-\omega_-^2\big)^2J_{++}
  +\alpha^2\big(J_{+-}+J_{-+}\big)
  +\big(\omega_+^2-\omega_c^2\big)^2J_{--}\Big]\Big\}, \label{eq: C_qq}\\[6pt]
  \langle Q_c(t)Q_c(t')\rangle\equiv C_{Q_cQ_c}(t,t')&=\alpha^2\!\int_0^{\min(t,t')}\!\!\!ds\;\Big[D_0\,\big(L_qg\big)(t-s)\,\big(L_qg\big)(t'-s)\notag\\
  &+D_2\;g(t-s)\,g(t'-s)\Big]\notag\\
  &=\frac{1}{\big(\omega_+^2-\omega_-^2\big)^2}
  \Big\{\alpha^2D_0\Big[\big(\omega_+^2-\omega_c^2\big)^2J_{++}
  +\alpha^2\big(J_{+-}+J_{-+}\big)
  +\big(\omega_c^2-\omega_-^2\big)^2J_{--}\Big]\notag\\
  &+\alpha^2D_2\Big[J_{++}-J_{+-}-J_{-+}+J_{--}\Big]\Big\}, \label{eq: C_QcQc}\\[6pt]
  \langle Q_c(t)q(t')\rangle\equiv C_{Q_cq}(t,t')&=-\,\alpha\!\int_0^{\min(t,t')}\!\!\!ds\;\Big[\alpha^2D_0\,\big(L_qg\big)(t-s)\,g(t'-s)\notag\\
  &+D_2\;g(t-s)\,\big(L_Qg\big)(t'-s)\Big]\notag\\
  &=\frac{-\alpha}{\big(\omega_+^2-\omega_-^2\big)^2}
  \Big\{\alpha^2D_0\Big[\big(\omega_+^2-\omega_c^2\big)\big(J_{+-}-J_{++}\big)
  +\big(\omega_c^2-\omega_-^2\big)\big(J_{--}-J_{-+}\big)\Big]\notag\\
  &+D_2\Big[\big(\omega_c^2-\omega_-^2\big)\big(J_{-+}-J_{++}\big)
  +\big(\omega_+^2-\omega_c^2\big)\big(J_{--}-J_{+-}\big)\Big]\Big\}. \label{eq: C_Qcq}
\end{align}

\subsection{The Brownian bridge}
\label{sec: Bridge}
One may, however, be interested instead in the conditioned propagator, where the final conditions for the classical field are specified, as for the Brownian bridge. We can thus obtain the probability of rare events (large deviations) as well as weak values and conditioned two-point functions (2pf). To this end, one generally writes the variables as $\phi=\phi^{\rm cl}+\delta\phi$, where $\phi^{\rm cl}$ satisfies the classical equations of motion. The propagator $Z(\phi_f,t_f;\phi_i,t_i)=\int \mathcal{D} \phi\, e^{-S}$ (with no integration over $\phi_f$) can then be rewritten as $Z=e^{-S_{\rm cl}}\int \mathcal{D} \delta\phi\,e^{-S[\delta\phi]}$, where $S_{\rm cl}$ is the semi-classical action. The factor $e^{-S_{\rm cl}}$ weights the final configurations, $\int d\phi_f\,Z=1$.

The normalised bridge two-point functions are then given by
\begin{align}
\langle\phi(t)\,\phi(s)\rangle_{\rm bridge}&=\phi^{\rm cl}(t)\,\phi^{\rm cl}(s)+\mathcal G(t,s), \label{eq: bridge_2pf}
\end{align}
where $\mathcal{G}$ obeys the boundary conditions $\mathcal{G}(t,s)|_{t_i}=\mathcal{G}(t,s)|_{t_f}=0$. Here $\phi^{\rm cl}(t)$ is the conditional mean at time $t$ and $\mathcal{G}$ is the conditional covariance, $\mathcal{G}(t,s)=\langle\phi(t)\,\phi(s)\rangle_{\rm bridge}-\langle\phi(t)\rangle\langle\phi(s)\rangle$. One may verify this by adding a current $J$ to the full action $S$:
\begin{align}
    Z[J](\phi_f,t_f;\phi_i,t_i)&=\int \mathcal{D}\phi\, e^{-S'}, \qquad
S'=S[\phi]+J\phi=S[\phi_{\rm cl}]+S[\delta\phi]+J(\delta\phi+\phi_{\rm cl}), \label{eq: Z_J}\\
\langle\phi(t)\phi(s)\rangle&=\frac{1}{Z[0]}\frac{\delta^2 Z[J]}{\delta J(t)\, \delta J(s)}\bigg|_{J=0}, \label{eq: 2pf_from_Z}
\end{align}
where $S[\phi]=S[\phi_{\rm cl}]+S[\delta\phi]$ is exact in the Gaussian case.

Alternatively, one may obtain the bridge covariance $\mathcal{G}(t,s)$ from the unconditional 2pf, using the law of total covariance,
\begin{align}
    \mathrm{Cov}(x,y)&=\mathbb{E}\big[\mathrm{Cov}(x,y|z)\big]+\mathrm{Cov}\big(\mathbb E[x|z],\mathbb E[y|z]\big). \label{eq: total_cov}
\end{align}
In the Gaussian case, $\mathbb E[\mathrm{Cov}(x,y|z)]=\mathrm{Cov}(x,y|z)$ and $\mathrm{Cov}(\mathbb E[x|z],\mathbb E[y|z])=\Sigma_{xz}\Sigma^{-1}_{zz}\Sigma_{zy}$, where $\Sigma$ is the covariance matrix. Then, conditioning on $\phi(t_f)$,
\begin{align}
    \mathcal{G}(t,s)&=\langle \phi(t)\phi(s)\rangle-\langle \phi(t)\phi(t_f)\rangle \langle \phi(t_f)\phi(t_f)\rangle^{-1}\langle \phi(t_f)\phi(s)\rangle. \label{eq: G_total_cov}
\end{align}
One can also show the equivalence of the two formulas for $\mathcal{G}$ from the relationship between the pinned partition function $Z$ and the free one, $Z_{\rm free}$, as we do in Appendix~\ref{app: 2pf}.

\begin{figure}[t]
  \centering
  \includegraphics[width=\linewidth]{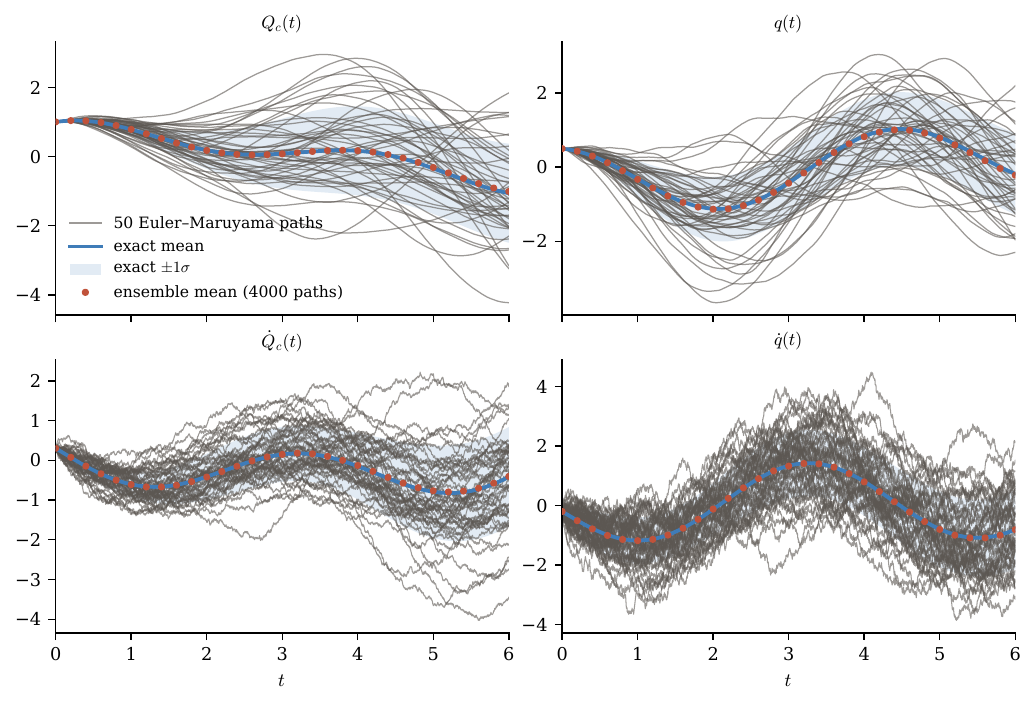}
  \caption{Euler--Maruyama integration of the Langevin equations~\eqref{eq: Langevin_Q} and~\eqref{eq: Langevin_q} with time step $\Delta t=10^{-3}$, for $\omega_Q=0.9$, $\omega_c=1.3$, $\alpha=0.7$, $D_0=0.25$, $D_2=1$ (so that $D_0D_2=1/4$, $\eta=1$), from the initial data $(Q_c,\dot Q_c,q,\dot q)=(1,0.3,0.5,-0.2)$. Grey: 50 sample paths. Blue: the exact mean, which coincides with the free on-shell solution of Fig.~\ref{fig: onshell}, and the exact $\pm1\sigma$ band. Red dots: the mean over 4000 paths. Top: positions; bottom: velocities.}
    \label{fig: langevin}
\end{figure}

Let us now turn back to the classical-quantum action. Varying it, and defining
\begin{align}
  P&\equiv iQ_q , \label{eq: P_def}
\end{align}
we obtain the equations of motion
\begin{align}
  \ddot Q_c+\omega_Q^2Q_c+\alpha q&=D_0\alpha^2P, \label{eq: eom_1}\\
  -\ddot P-\omega_Q^2P&=\alpha D_2^{-1}\big(\ddot q+\omega_c^2q+\alpha Q_c\big), \label{eq: eom_2}\\
  \big(\partial_t^2+\omega_c^2\big)\big(\ddot q+\omega_c^2q+\alpha Q_c\big)&=-\alpha D_2P . \label{eq: eom_3}
\end{align}
Eliminating $P$ using the first equation gives the action
\begin{align}
  S_{\rm sc}[Q_c,q]&= \Big[\frac{\dot Q_c\big(\ddot Q_c+\omega_Q^2Q_c+\alpha q\big)}{D_0\alpha^{2}}\Big]_{t_i}^{t_f}\notag\\
  &-\int_{t_i}^{t_f}\!dt\;\Big[\frac{\big(\ddot Q_c+\omega_Q^2Q_c+\alpha q\big)^{2}}{2D_0\alpha^{2}}
  +\frac{D_2^{-1}}{2}\big(\ddot q+\omega_c^2q+\alpha Q_c\big)^{2}\Big]. \label{eq: S_sc}
\end{align}
This is the Onsager--Machlup action, with weight $e^{S_{\rm sc}}$, of two coupled harmonic oscillators
obeying the stochastic differential equations
\begin{align}
  \ddot Q_c+\omega_Q^2Q_c+\alpha q&=\xi_Q, \qquad\langle\xi_Q(t)\xi_Q(t')\rangle=\alpha^2D_0\,\delta(t-t'), \label{eq: Langevin_Q}\\
  \ddot q+\omega_c^2q+\alpha Q_c&=\xi_q, \qquad\langle\xi_q(t)\xi_q(t')\rangle=D_2\,\delta(t-t'). \label{eq: Langevin_q}
\end{align}
Sample paths of these equations are shown in Fig.~\ref{fig: langevin}. Since the noise acts as a force, $Q_c$ and $q$ are once differentiable, while their velocities are rough.
By the first equation of motion, the boundary term is
\begin{align}
  \Big[\frac{\dot Q_c\big(\ddot Q_c+\omega_Q^2Q_c+\alpha q\big)}{D_0\alpha^{2}}\Big]_{t_i}^{t_f}&=\big[\dot Q_c\,P\big]_{t_i}^{t_f}. \label{eq: S_sc_boundary}
\end{align}
Its final-time part vanishes because $P(t_f)=0$; its initial-time part vanishes for $\dot Q_c(t_i)=0$.

Inserting the equations of motion into the action leaves only boundary terms:
\begin{align}
  S_{\rm on\text{-}shell}&=\frac12\Big[\dot Q_c\,P+Q_c\,\dot P
  +D_2^{-1}\Big(q\big(\dddot q+\omega_c^2\dot q+\alpha\dot Q_c\big)
  -\dot q\big(\ddot q+\omega_c^2q+\alpha Q_c\big)\Big)\Big]_{t_i}^{t_f}. \label{eq: S_onshell_1}
\end{align}
The second equation of motion gives
$D_2^{-1}\big(\ddot q+\omega_c^2q+\alpha Q_c\big)=-\alpha^{-1}\big(\ddot P+\omega_Q^2P\big)$,
and its time derivative gives
$D_2^{-1}\big(\dddot q+\omega_c^2\dot q+\alpha\dot Q_c\big)=-\alpha^{-1}\big(\dddot P+\omega_Q^2\dot P\big)$.
In terms of $P$ alone, the on-shell action is therefore
\begin{align}
  S_{\rm on\text{-}shell}&=\frac12\Big[\dot Q_c\,P+Q_c\,\dot P
  -\frac{1}{\alpha}\Big(q\big(\dddot P+\omega_Q^2\dot P\big)-\dot q\big(\ddot P+\omega_Q^2P\big)\Big)\Big]_{t_i}^{t_f}, \label{eq: S_onshell_2}
\end{align}
subject to the boundary conditions
\begin{align}
  q(t_i)&=q_i,\quad \dot q(t_i)=\dot q_i,\quad q(t_f)=q_f,\quad \dot q(t_f)=\dot q_f,\qquad
  Q_c(t_i)=Q_i,\quad \dot Q_c(t_i)=\dot Q_i,\qquad P(t_f)=\dot P(t_f)=0 . \label{eq: bridge_bcs}
\end{align}
Using $P(t_f)=\dot P(t_f)=0$, this evaluates to
\begin{align}
  S_{\rm on\text{-}shell}&=-\frac{1}{2\alpha}\Big[q_f\,\dddot P(t_f)-\dot q_f\,\ddot P(t_f)\Big]
  -\frac12\Big[\dot Q_i\,P(t_i)+Q_i\,\dot P(t_i)\Big]\notag\\
  &+\frac{1}{2\alpha}\Big[q_i\big(\dddot P+\omega_Q^2\dot P\big)(t_i)-\dot q_i\big(\ddot P+\omega_Q^2P\big)(t_i)\Big]. \label{eq: S_onshell_3}
\end{align}

In the free problem the on-shell action vanishes. There, $P=0$ everywhere, and the conditions at $t_f$
are the natural ones,
\begin{align}
  \ddot q+\omega_c^2q+\alpha Q_c&=\dddot q+\omega_c^2\dot q+\alpha\dot Q_c=0
  \quad\Longleftrightarrow\quad
  \ddot P(t_f)=\dddot P(t_f)=0 \qquad\text{at } t=t_f . \label{eq: natural_bcs}
\end{align}
In the bridge, $q_f$ and $\dot q_f$ are prescribed instead, so the $t_f$ term survives.

We have thus rewritten the on-shell action in terms of the boundary conditions and $P$. All that remains is to solve for $P$ and, by extension, for $q$ and $Q_c$, to obtain the conditioned expectation values of $q$ and $Q^{\pm}$. These are called weak values \cite{aharonov1988result}: $\langle Q^{\pm}\rangle_{\rm bridge}=\operatorname{Tr}\big[\hat E\,\hat Q\hat\rho\big]/\operatorname{Tr}\big[\hat E\hat\rho\big]$, where conditioning on final data inserts a post-selected effect operator $\hat E$ at $t_f$.
Note that these weak values, although complex for the quantum fields, are by definition \emph{not} anomalous \cite{PhysRevLett.113.200401}, given that the quantum system can be described by a Gaussian probability distribution.

\begin{figure}[t]
  \centering
  \includegraphics[width=\linewidth]{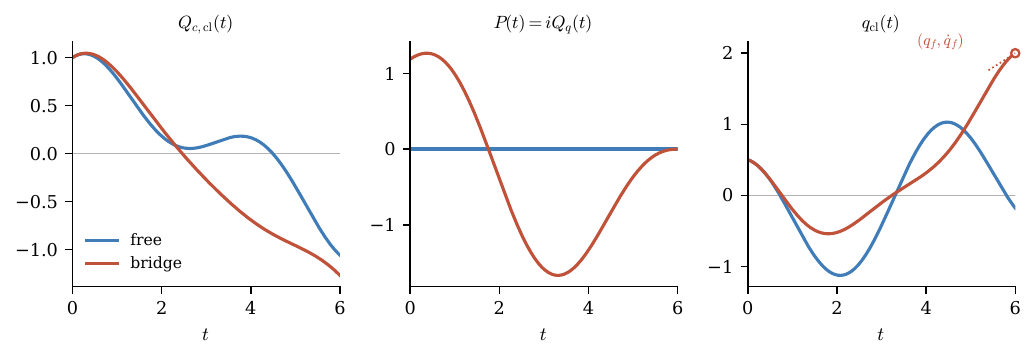}
  \caption{Solutions of the equations of motion~\eqref{eq: eom_1}--\eqref{eq: eom_3} for the parameters and initial data of Fig.~\ref{fig: langevin}, with $t_f=6$. Blue: the free problem, with natural boundary conditions at $t_f$, for which $P\equiv0$. Red: the bridge, with $q(t_f)=2$ and $\dot q(t_f)=0.4$ prescribed (open circle and dotted slope) and $P(t_f)=\dot P(t_f)=0$. The bridge solution coincides with the conditional mean $\Phi_{\rm det}+\mathbf h\Sigma^{-1}\delta Y$ of Eq.~\eqref{eq: equiv_J1}.}
  \label{fig: onshell}
\end{figure}

Solving the equations of motion, we obtain (see Appendix~\ref{app: EOM} for the calculational details)
\begin{align}
P(t)&=c_1\cos\omega_+t+c_2\sin\omega_+t+c_3\cos\omega_-t+c_4\sin\omega_-t , \label{eq: P_sol}\\[6pt]
  Q_c(t)&=\frac{1}{2(\omega_+^2-\omega_-^2)}\Big\{t\,\big[\alpha^2D_0(\omega_c^2-\omega_+^2)+D_2(\omega_Q^2-\omega_+^2)\big]\frac{-c_1\sin\omega_+t+c_2\cos\omega_+t}{\omega_+}\notag\\
  &-2(\alpha^2D_0+D_2)\big(c_1\cos\omega_+t+c_2\sin\omega_+t\big)\notag\\
  &-t\,\big[\alpha^2D_0(\omega_c^2-\omega_-^2)+D_2(\omega_Q^2-\omega_-^2)\big]\frac{-c_3\sin\omega_-t+c_4\cos\omega_-t}{\omega_-}\notag\\
  &+2(\alpha^2D_0+D_2)\big(c_3\cos\omega_-t+c_4\sin\omega_-t\big)\Big\}\notag\\
  &+c_5\cos\omega_+t+c_6\sin\omega_+t+c_7\cos\omega_-t+c_8\sin\omega_-t, \label{eq: Qc_sol}\\[6pt]
    q(t)&=-\frac{1}{2\alpha(\omega_+^2-\omega_-^2)}\Big\{t\,\big[D_2(\omega_Q^2-\omega_+^2)^2+\alpha^4D_0\big]\frac{-c_1\sin\omega_+t+c_2\cos\omega_+t}{\omega_+}\notag\\
  &-4D_2(\omega_Q^2-\omega_+^2)\big(c_1\cos\omega_+t+c_2\sin\omega_+t\big)\notag\\
  &-t\,\big[D_2(\omega_Q^2-\omega_-^2)^2+\alpha^4D_0\big]\frac{-c_3\sin\omega_-t+c_4\cos\omega_-t}{\omega_-}\notag\\
  &+4D_2(\omega_Q^2-\omega_-^2)\big(c_3\cos\omega_-t+c_4\sin\omega_-t\big)\Big\}\notag\\
  &+\frac{\omega_+^2-\omega_Q^2}{\alpha}\big(c_5\cos\omega_+t+c_6\sin\omega_+t\big)
  +\frac{\omega_-^2-\omega_Q^2}{\alpha}\big(c_7\cos\omega_-t+c_8\sin\omega_-t\big), \label{eq: q_sol}
\end{align}
with $c_1,\dots,c_8$ functions of the boundary data. The free and bridge solutions are compared in Fig.~\ref{fig: onshell}.

\emph{Critical and supercritical cases.}
The solution above assumes $\omega_-^2>0$.

At the critical coupling $\alpha^2=\omega_Q^2\omega_c^2$ we have $\omega_-=0$ and
$\omega_+^2=\omega_Q^2+\omega_c^2$. The terms carrying $1/\omega_-$ are then singular, so the basis
has to be replaced rather than taken to the limit term by term. The operator becomes
$(\partial_t^2+\omega_+^2)\,\partial_t^2$, and its kernel is spanned by
$\cos\omega_+t$, $\sin\omega_+t$, $1$ and $t$. One therefore writes
\begin{align}
  P(t)&=c_1\cos\omega_+t+c_2\sin\omega_+t+c_3+c_4\,t . \label{eq: P_sol_crit}
\end{align}
The resonant solution of $(\partial_t^2+\omega_+^2)\partial_t^2Y=P$ is
\begin{align}
  Y&=\frac{t}{2\omega_+^4}\big(-c_1\omega_+\sin\omega_+t+c_2\omega_+\cos\omega_+t\big)
  +\frac{1}{\omega_+^2}\Big(\frac{c_3t^2}{2}+\frac{c_4t^3}{6}\Big). \label{eq: Y_crit}
\end{align}
The homogeneous part of $(Q_c,q)$ becomes
\begin{align}
  Q_c&\supset c_5\cos\omega_+t+c_6\sin\omega_+t+c_7+c_8\,t,
  \qquad
  q\supset \frac{\omega_c^2}{\alpha}\big(c_5\cos\omega_+t+c_6\sin\omega_+t\big)-\frac{\omega_Q^2}{\alpha}\big(c_7+c_8\,t\big). \label{eq: hom_crit}
\end{align}
$Q_c$ and $q$ follow from $Y$ through the same two operators as before, and the eight conditions are
imposed in the same way. The growth is now polynomial: $P$ is at most linear in $t$, while
$Q_c$ and $q$ grow up to $t^3$.

In the supercritical case, $\alpha^2>\omega_Q^2\omega_c^2$, we set $\omega_-^2=-\kappa^2<0$. The formulas
carry over unchanged under $\omega_-\to i\kappa$, with $c_3$ and $c_4$ redefined so that they stay real:
\begin{align}
  \cos\omega_-t&\to\cosh\kappa t,\qquad \frac{\sin\omega_-t}{\omega_-}\to\frac{\sinh\kappa t}{\kappa},\qquad
  \omega_+^2-\omega_-^2\to\omega_+^2+\kappa^2 . \label{eq: supercrit_map}
\end{align}
$P$ then grows like $e^{\kappa t}$, and $Q_c$ and $q$ pick up resonant terms of order $t\,e^{\kappa t}$.
The growth is exponential, consistent with the exponential growth of all the two-point functions
in this regime.

To find the conditioned covariance in our case, $\mathcal{G}_{XY}=(\mathcal{M}_{\rm pinned}^{-1})_{XY}$, we invert
\begin{align}
  \mathcal{M}_{\rm pinned}&=  \begin{bmatrix}
    \frac{\alpha^2}{D_2}&iL_Q^{A}&\frac{\alpha}{D_2}L_q^R \\iL_Q^R&\alpha^2D_0&i\alpha\\ \frac{\alpha}{D_2}L_q^A&i\alpha&\frac{1}{D_2}(L_q  L_q)_{CC}
    \end{bmatrix}, \label{eq: M_pinned}
\end{align}
where
\begin{align}
    \mathcal{I}_{CQ}(\delta\vec{\phi})&=-\frac{1}{2}\int^{t_f}_{t_i}dt\, \delta\vec{\phi}^{T}\mathcal{M}_{\rm pinned}\delta\vec{\phi}, \label{eq: action_pinned}
\end{align}
with $\vec{\phi}$ and $L_{Q/q}^{R/A}$ as in Eq.~\eqref{eq: M_matrix}, and $(L_q  L_q)_{CC}=(\partial_t^2+\omega^2_{c})^2$ with clamped-clamped boundary conditions.

We could directly invert this matrix to find the conditional covariance; however, given that we already have the free correlators, it is more convenient to use the relation
\begin{align}
    \langle \phi(t)\phi(s)\rangle_{{\rm conn}, XY}^{\rm bridge}\equiv \mathcal{G}_{XY}&=    \langle \phi(t)\phi(s)\rangle_{XY}-    \mathbf h_X(t)\,  \Sigma^{-1}\mathbf h_Y^{\mathsf T} (s), \label{eq: G_XY}\\
    \Sigma&=\begin{pmatrix}
  \langle q(t_f)\,q(t_f)\rangle & \langle q(t_f)\,\dot q(t_f)\rangle\\[4pt]
  \langle \dot q(t_f)\,q(t_f)\rangle & \langle \dot q(t_f)\,\dot q(t_f)\rangle
  \end{pmatrix},\qquad \mathbf h_X(t)=\begin{pmatrix}
      \langle\phi_X(t)\, q(t_f)\rangle\\\langle\phi_X(t)\, \dot q(t_f)\rangle
  \end{pmatrix}, \label{eq: Sigma_h}
\end{align}

Let us now compute $\Sigma^{-1}$. Using the expression for $\langle q(t)q(t')\rangle$, we have
\begin{align}
  \langle q(t_f)\,q(t_f)\rangle&=\int_0^{T}\!d\tau\,\Big[\alpha^4D_0\,g(\tau)^2+D_2\,\big(L_Qg\big)(\tau)^2\Big], \label{eq: Sigma_11}\\
  \langle \dot q(t_f)\,\dot q(t_f)\rangle&=\int_0^{T}\!d\tau\,\Big[\alpha^4D_0\,\dot g(\tau)^2+D_2\,\big(\partial_\tau L_Qg\big)(\tau)^2\Big], \label{eq: Sigma_22}\\
  \langle q(t_f)\,\dot q(t_f)\rangle&=\int_0^{T}\!d\tau\,\Big[\alpha^4D_0\,g(\tau)\dot g(\tau)+D_2\,\big(L_Qg\big)(\tau)\big(\partial_\tau L_Qg\big)(\tau)\Big]
  =\frac12\,\frac{d}{dt_f}\,\langle q(t_f)\,q(t_f)\rangle . \label{eq: Sigma_12}
\end{align}
The last equality holds because the integrand is $\tfrac12\partial_\tau$ of the integrand of
$\langle q(t_f)q(t_f)\rangle$ and the integral starts at $g(0)=(L_Qg)(0)=0$.

Then,
\begin{align}
  \det\Sigma&=\langle q(t_f)\,q(t_f)\rangle\,\langle \dot q(t_f)\,\dot q(t_f)\rangle-\langle q(t_f)\,\dot q(t_f)\rangle^2 , \label{eq: detSigma}\\
  \Sigma^{-1}&=\frac{1}{\det\Sigma}
  \begin{pmatrix}
  \langle \dot q(t_f)\,\dot q(t_f)\rangle & -\langle q(t_f)\,\dot q(t_f)\rangle\\[4pt]
  -\langle q(t_f)\,\dot q(t_f)\rangle & \langle q(t_f)\,q(t_f)\rangle
  \end{pmatrix}, \label{eq: Sigma_inv}
\end{align}
with $\det\Sigma>0$ for every $t_f>0$.
For small $t_f$,
\begin{align}
  \det\Sigma&\simeq\frac{D_2^2\,t_f^4}{12}. \label{eq: detSigma_smallT}
\end{align}

\emph{Bridge covariances ($t,t'<t_f$).}
\begin{align}
  \big\langle q(t)q(t')\big\rangle^{\rm bridge}
  &=C_{qq}(t,t')-\frac{1}{\det\Sigma}\Big[\Sigma_{22}\,C_{qq}(t,t_f)\,C_{qq}(t',t_f)\notag\\
  &-\Sigma_{12}\Big(C_{qq}(t,t_f)\,\partial_{t_f}C_{qq}(t',t_f)+\partial_{t_f}C_{qq}(t,t_f)\,C_{qq}(t',t_f)\Big)\notag\\
  &+\Sigma_{11}\,\partial_{t_f}C_{qq}(t,t_f)\,\partial_{t_f}C_{qq}(t',t_f)\Big], \label{eq: bridge_qq}
  \\[6pt]
  \big\langle Q_c(t)Q_c(t')\big\rangle^{\rm bridge}
  &=C_{Q_cQ_c}(t,t')-\frac{1}{\det\Sigma}\Big[\Sigma_{22}\,C_{Q_cq}(t,t_f)\,C_{Q_cq}(t',t_f)\notag\\
  &-\Sigma_{12}\Big(C_{Q_cq}(t,t_f)\,\partial_{t_f}C_{Q_cq}(t',t_f)+\partial_{t_f}C_{Q_cq}(t,t_f)\,C_{Q_cq}(t',t_f)\Big)\notag\\
  &+\Sigma_{11}\,\partial_{t_f}C_{Q_cq}(t,t_f)\,\partial_{t_f}C_{Q_cq}(t',t_f)\Big], \label{eq: bridge_QcQc}
  \\[6pt]
  \big\langle Q_c(t)q(t')\big\rangle^{\rm bridge}
  &=C_{Q_cq}(t,t')-\frac{1}{\det\Sigma}\Big[\Sigma_{22}\,C_{Q_cq}(t,t_f)\,C_{qq}(t',t_f)\notag\\
  &-\Sigma_{12}\Big(C_{Q_cq}(t,t_f)\,\partial_{t_f}C_{qq}(t',t_f)+\partial_{t_f}C_{Q_cq}(t,t_f)\,C_{qq}(t',t_f)\Big)\notag\\
  &+\Sigma_{11}\,\partial_{t_f}C_{Q_cq}(t,t_f)\,\partial_{t_f}C_{qq}(t',t_f)\Big], \label{eq: bridge_Qcq}
  \\[6pt]
  \big\langle Q_q(t)Q_q(t')\big\rangle^{\rm bridge}
  &=\frac{\alpha^2}{\det\Sigma}\Big[\Sigma_{22}\,g(t_f-t)\,g(t_f-t')\notag\\
  &-\Sigma_{12}\Big(g(t_f-t)\,\dot g(t_f-t')+\dot g(t_f-t)\,g(t_f-t')\Big)\notag\\
  &+\Sigma_{11}\,\dot g(t_f-t)\,\dot g(t_f-t')\Big], \label{eq: bridge_QqQq}
  \\[6pt]
  \big\langle q(t)Q_q(t')\big\rangle^{\rm bridge}
  &=i\alpha\Bigg\{g(t-t')-\frac{1}{\det\Sigma}\Big[\Big(\Sigma_{22}\,C_{qq}(t,t_f)-\Sigma_{12}\,\partial_{t_f}C_{qq}(t,t_f)\Big)g(t_f-t')\notag\\
  &+\Big(\Sigma_{11}\,\partial_{t_f}C_{qq}(t,t_f)-\Sigma_{12}\,C_{qq}(t,t_f)\Big)\dot g(t_f-t')\Big]\Bigg\}, \label{eq: bridge_qQq}
  \\[6pt]
  \big\langle Q_c(t)Q_q(t')\big\rangle^{\rm bridge}
  &=-i\Bigg\{(L_qg)(t-t')+\frac{\alpha}{\det\Sigma}\Big[\Big(\Sigma_{22}\,C_{Q_cq}(t,t_f)-\Sigma_{12}\,\partial_{t_f}C_{Q_cq}(t,t_f)\Big)g(t_f-t')\notag\\
  &+\Big(\Sigma_{11}\,\partial_{t_f}C_{Q_cq}(t,t_f)-\Sigma_{12}\,C_{Q_cq}(t,t_f)\Big)\dot g(t_f-t')\Big]\Bigg\}. \label{eq: bridge_QcQq}
\end{align}
Note that, unlike in the free case, $\langle Q_q Q_q \rangle^{\rm bridge}\neq 0$. The original variables $Q^\pm$ have both a complex mean and complex bridge two-point functions.

\subsection{Energy and heating from two-point functions}
\label{sec: Heating}
We can now compute the expectation (weak) value of the Hamiltonian $H$ in the free (bridge) case. The energy can be inserted on either branch:
\begin{align}
  H^\pm&=\tfrac12\big(\dot Q^{\pm\,2}+\dot q^2\big)+\tfrac12\big(\omega_Q^2Q^{\pm\,2}+\omega_c^2q^2\big)+\alpha\,Q^\pm q . \label{eq: H_pm}
\end{align}
Expanding,
\begin{align}
  H^\pm&=H_c\;\pm\;\tfrac12\Big(\dot Q_c\dot Q_q+\omega_Q^2\,Q_cQ_q+\alpha\,q\,Q_q\Big)
  \;+\;\tfrac18\Big(\dot Q_q^{\,2}+\omega_Q^2\,Q_q^2\Big), \label{eq: H_pm_expanded}
\end{align}
where $H_c$ is $H$ evaluated on $(Q_c,q)$. We define the two equal-time combinations
\begin{align}
  \Delta_1(t)&=\big\langle\dot Q_c\dot Q_q\big\rangle+\omega_Q^2\big\langle Q_cQ_q\big\rangle+\alpha\big\langle qQ_q\big\rangle,
  \qquad
  \Delta_2(t)=\big\langle\dot Q_q\dot Q_q\big\rangle+\omega_Q^2\big\langle Q_qQ_q\big\rangle . \label{eq: Delta12}
\end{align}
The possible definitions of the energy are then
\begin{align}
  \big\langle H^\pm\big\rangle&=\big\langle H_c\big\rangle\pm\tfrac12\Delta_1+\tfrac18\Delta_2,
  \qquad
  \tfrac12\big\langle H^++H^-\big\rangle=\big\langle H_c\big\rangle+\tfrac18\Delta_2 . \label{eq: H_orderings}
\end{align}
These correspond to $\operatorname{Tr}[\hat H\hat\rho]$, $\operatorname{Tr}[\hat\rho\hat H]$, and the symmetric ordering, respectively.

\emph{Free case: no ambiguity.}
There should be no ambiguity in the free case, given that
\begin{align}
  \operatorname{Tr}\big[\hat H\hat\rho(t)\big]&=\operatorname{Tr}\big[\hat\rho(t)\hat H\big], \label{eq: trace_cyclic}
\end{align}
and one can indeed show this by working in phase space: the equal-time $\langle P_cP_q\rangle$ is proportional to $\theta(0)\,[\hat P,\hat P]=0$, while $\langle P_q P_q\rangle$ vanishes (as does $\langle Q_q Q_q\rangle$). However, since velocities are used instead of momenta, we pick up a spurious contact term if we use $H^{\pm}$ instead of $\frac{1}{2}(H^++H^-)$:
\begin{align}
  \langle\dot Q_c(t)\dot Q_q(t)\rangle&=i\,\partial_\tau(L_qg)(0)\,\delta(0)=i\,\delta(0),
  \qquad
  \big\langle H^\pm\big\rangle=\big\langle H_c\big\rangle\pm\tfrac{i}{2}\,\delta(0). \label{eq: contact_term}
\end{align}
In velocity form, the energy is therefore $H_c=H(Q_c,\dot Q_c,q,\dot q)$.

\emph{The bridge case.}
Conditioning on final data inserts a post-selected effect operator $\hat E$ at $t_f$, and
$[\hat E,\hat H]\neq0$. The two branches give different, complex-conjugate weak values:
\begin{align}
  \big\langle H^+\big\rangle^{\rm bridge}&=\frac{\operatorname{Tr}\big[\hat E\,\hat H\hat\rho\big]}{\operatorname{Tr}\big[\hat E\hat\rho\big]},
  \qquad
  \big\langle H^-\big\rangle^{\rm bridge}=\frac{\operatorname{Tr}\big[\hat E\,\hat\rho\hat H\big]}{\operatorname{Tr}\big[\hat E\hat\rho\big]}. \label{eq: H_weak}
\end{align}
They differ by the finite part of $\Delta_1$ as well as by the contact term: unlike in the unconditioned case, the difference is not an artifact of working in position space. The real, finite energy is the real part
of the weak value, i.e.,\ the symmetric insertion:
\begin{align}
  \big\langle H\big\rangle^{\rm bridge}
  &=\frac{\operatorname{Tr}\big[\hat E\,\tfrac12\{\hat H,\hat\rho\}\big]}{\operatorname{Tr}\big[\hat E\hat\rho\big]}
  =\tfrac12\big\langle H^++H^-\big\rangle^{\rm bridge}
  =\big\langle H_c\big\rangle^{\rm bridge}
  +\tfrac18\Big(\big\langle\dot Q_q\dot Q_q\big\rangle+\omega_Q^2\big\langle Q_qQ_q\big\rangle\Big)^{\rm bridge}. \label{eq: H_bridge_sym}
\end{align}
In this combination, both $\pm\tfrac{i}{2}\delta(0)$ and the imaginary $\Delta_1$ cancel.
The last term is the $\hbar^2$ Moyal correction. It vanishes in the unconditional problem.
In the bridge it is nonzero, because there
$\langle Q_q(t)Q_q(s)\big\rangle^{\rm bridge}\neq0$.

Having settled the ordering ambiguities, we can compute the expectation value of the Hamiltonian. The connected part of $\langle H(t)\rangle$, i.e., the part independent of the initial state, is the coincident limit of the two-point
functions:
\begin{align}
  \big\langle H(t)\big\rangle_{\rm conn}
  &=\tfrac12\Big[\partial_t\partial_{t'}C_{Q_cQ_c}+\partial_t\partial_{t'}C_{qq}
  +\omega_Q^2C_{Q_cQ_c}+\omega_c^2C_{qq}+2\alpha\,C_{Q_cq}\Big]_{t'=t}. \label{eq: H_conn}
\end{align}

One can compute this directly using the correlation functions~\eqref{eq: C_qq}--\eqref{eq: C_Qcq}, or using the Langevin equations~\eqref{eq: Langevin_Q} and~\eqref{eq: Langevin_q} obtained from the action, with $\langle\xi_Q\xi_q\rangle=0$. Using the latter, we obtain
\begin{align}
  \frac{dH}{dt}&=\dot Q_c\,\xi_Q+\dot q\,\xi_q . \label{eq: dHdt}
\end{align}
The solution is
\begin{align}
  Q_c&=Q_{c,\rm det}+(L_qg) \circ \xi_Q-\alpha\,g\circ\xi_q,\qquad
  q=q_{\rm det}+(L_Qg)\circ\xi_q-\alpha\,g\circ\xi_Q, \label{eq: Langevin_sol}
\end{align}
with $(k\circ\xi)(t)=\int_0^tk(t-s)\,\xi(s)\,ds$, where $Q_{c,\rm det}$ and $q_{\rm det}$ are the solutions of the noiseless evolution, and $L_Q$, $L_q$ are as in Eq.~\eqref{eq: M_matrix}.
Since $g(0)=(L_qg)(0)=(L_Qg)(0)=0$, the velocities are
\begin{align}
  \dot Q_c&=\dot Q_{c,\rm det}+\partial_\tau(L_qg)\circ\xi_Q-\alpha\,\dot g\circ\xi_q,\qquad
  \dot q=\dot q_{\rm det}+\partial_\tau(L_Qg)\circ\xi_q-\alpha\,\dot g\circ\xi_Q . \label{eq: Langevin_vel}
\end{align}
The noiseless parts are uncorrelated with the noise, and the two noises are independent.
Hence
\begin{align}
  \big\langle\dot Q_c(t)\,\xi_Q(t)\big\rangle&=\alpha^2D_0\int_0^t\!ds\;\partial_\tau(L_qg)(t-s)\,\delta(t-s)
  =\tfrac12\,\alpha^2D_0\,\partial_\tau(L_qg)(0)=\tfrac12\,\alpha^2D_0 , \label{eq: Qdot_xi}\\
  \big\langle\dot q(t)\,\xi_q(t)\big\rangle&=\tfrac12\,D_2\,\partial_\tau(L_Qg)(0)=\tfrac12\,D_2 , \label{eq: qdot_xi}
\end{align}
where we use the endpoint half-weight
$\int_0^t\delta(t-s)\,ds=\tfrac12$. Therefore
\begin{align}
  \frac{d}{dt}\big\langle H\big\rangle&=\tfrac12\big(\alpha^2D_0+D_2\big) \label{eq: heating_rate}
\end{align}
and
\begin{align}
  \big\langle H(t)\big\rangle&=H\big[X_{\rm det}(0)\big]+\tfrac12\big(\alpha^2D_0+D_2\big)\,t. \label{eq: H_free}
\end{align}

\emph{Bridge rate.} For the bridge, we again use Eq.~\eqref{eq: dHdt} for $H_c$.
Our goal will once more be to obtain $\langle\dot Q_c\,\xi_Q\rangle$ and $\langle\dot q\,\xi_q\rangle$ from the action, written in terms of the noise:
\begin{align}
  S&=-\int_0^{t_f}\!dt\,\Big[\frac{\xi_q^2}{2D_2}+iQ_q\,\xi_Q+\frac{\alpha^2D_0}{2}\,Q_q^2\Big]. \label{eq: S_noise}
\end{align}
We use the fact that, for any interior time $0<t<t_f$, the integral of a total functional derivative
vanishes:
\begin{align}
  \int\!\mathcal D\Phi\;\frac{\delta}{\delta\Phi(t)}\Big(X\,e^{S}\Big)&=0
  \quad\Longrightarrow\quad
  \Big\langle\frac{\delta X}{\delta\Phi(t)}\Big\rangle+\Big\langle X\,\frac{\delta S}{\delta\Phi(t)}\Big\rangle=0 . \label{eq: SD_identity}
\end{align}
We vary the action with respect to $Q_q$ and $Q_c$, so that $\big\langle\frac{\delta X}{\delta\Phi(t)}\big\rangle=0$ for our desired expectation values. Here $\delta S/\delta Q_q=-i\,\xi_Q-\alpha^2D_0\,Q_q$ and $\delta S/\delta Q_c=-\frac{\alpha}{D_2}\xi_q-i\big(\ddot Q_q+\omega_Q^2Q_q\big)$. Using the linearity of the expectation value, we obtain
\begin{align}
  \big\langle X\,\xi_Q(t)\big\rangle&=i\alpha^2 D_0\,\big\langle X\,Q_q(t)\big\rangle , \label{eq: X_xiQ}\\
  \big\langle X\,\xi_q(t)\big\rangle&=-\frac{iD_2}{\alpha}\,\big\langle X\,\big(\ddot Q_q+\omega_Q^2Q_q\big)(t)\big\rangle . \label{eq: X_xiq}
\end{align}
The bridge subtraction is given by
\begin{align}
  \big\langle\dot Q_c(t)Q_q(t)\big\rangle_{\rm sub}
  &=-\frac{i\alpha}{\det\Sigma}\Big[\big(\Sigma_{22}\partial_tC_{Q_cq}(t,t_f)-\Sigma_{12}\partial_t\partial_{t_f}C_{Q_cq}(t,t_f)\big)g\notag\\
  &+\big(\Sigma_{11}\partial_t\partial_{t_f}C_{Q_cq}(t,t_f)-\Sigma_{12}\partial_tC_{Q_cq}(t,t_f)\big)\dot g\Big], \label{eq: sub_Qc}\\
  \big\langle\dot q(t)\big(\ddot Q_q+\omega_Q^2Q_q\big)(t)\big\rangle_{\rm sub}
  &=-\frac{i\alpha}{\det\Sigma}\Big[\big(\Sigma_{22}\partial_tC_{qq}(t,t_f)-\Sigma_{12}\partial_t\partial_{t_f}C_{qq}(t,t_f)\big)L_Qg\notag\\
  &+\big(\Sigma_{11}\partial_t\partial_{t_f}C_{qq}(t,t_f)-\Sigma_{12}\partial_tC_{qq}(t,t_f)\big)\partial_\tau L_Qg\Big]. \label{eq: sub_q}
\end{align}
Therefore,
\begin{align}
  \frac{d}{dt}\big\langle H_c\big\rangle^{\rm bridge}_{\rm conn}
  &=\tfrac12\big(\alpha^2D_0+D_2\big)-\frac{1}{\det\Sigma}\Big\{
  \Sigma_{22}\Big[-\alpha^3D_0\,g(t_f-t)\,\partial_tC_{Q_cq}(t,t_f)+D_2\,(L_Qg)(t_f-t)\,\partial_tC_{qq}(t,t_f)\Big]\notag\\
  &-\Sigma_{12}\Big[-\alpha^3D_0\,\dot g(t_f-t)\,\partial_tC_{Q_cq}(t,t_f)+D_2\,(\partial_\tau L_Qg)(t_f-t)\,\partial_tC_{qq}(t,t_f)\notag\\
  &-\alpha^3D_0\,g(t_f-t)\,\partial_t\partial_{t_f}C_{Q_cq}(t,t_f)+D_2\,(L_Qg)(t_f-t)\,\partial_t\partial_{t_f}C_{qq}(t,t_f)\Big]\notag\\
  &+\Sigma_{11}\Big[-\alpha^3D_0\,\dot g(t_f-t)\,\partial_t\partial_{t_f}C_{Q_cq}(t,t_f)+D_2\,(\partial_\tau
  L_Qg)(t_f-t)\,\partial_t\partial_{t_f}C_{qq}(t,t_f)\Big]\Big\}. \label{eq: bridge_rate}
\end{align}
Here
\begin{align}
  \Sigma_{11}&=C_{qq}(t_f,t_f),\qquad
  \Sigma_{12}=\partial_{t'}C_{qq}(t_f,t')\big|_{t'=t_f},\qquad
  \Sigma_{22}=\partial_t\partial_{t'}C_{qq}\big|_{t=t'=t_f}, \label{eq: Sigma_entries}
\end{align}
and every $C$ is the $J_{ab}$ combination given in Eqs.~\eqref{eq: C_qq}--\eqref{eq: C_Qcq}. This is the rate of heating of $H_c$ attributable solely to fluctuations, independent of the initial and final states.

\begin{figure}[t]
  \centering
  \includegraphics[width=\linewidth]{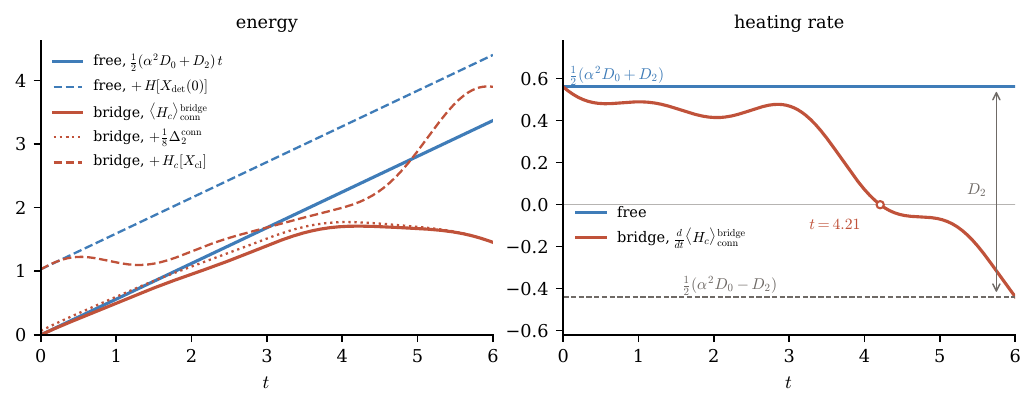}
  \caption{Energy of the pair in the free (blue) and bridge (red) cases, for the parameters and initial data of Fig.~\ref{fig: langevin}, with $t_f=6$. Left: the solid lines are the connected, initial-state-independent parts, $\tfrac12(\alpha^2D_0+D_2)t$ [Eq.~\eqref{eq: H_free}] and $\langle H_c\rangle^{\rm bridge}_{\rm conn}$; the dotted line adds the Moyal term $\tfrac18\Delta_2^{\rm conn}$ of Eq.~\eqref{eq: Delta2_conn}. The dashed lines add the contribution of the initial data and, for the bridge, of the record $(q_f,\dot q_f)=(2,0.4)$: $H[X_{\rm det}(0)]$ and $H_c[X_{\rm cl}]$, respectively. Right: heating rates. The free rate~\eqref{eq: heating_rate} is constant; the bridge rate~\eqref{eq: bridge_rate} starts at the free value, changes sign at $t\simeq4.21$, and reaches $\tfrac12(\alpha^2D_0-D_2)$ at $t_f$, Eq.~\eqref{eq: bridge_rate_tf}.}
  \label{fig: energy}
\end{figure}

At $t\to0^+$, all $\partial_tC_{Xq}(t,t_f)$ and $\partial_t\partial_{t_f}C_{Xq}(t,t_f)$ vanish, so the bridge
rate equals the free rate $\tfrac12(\alpha^2D_0+D_2)$.
At $t\to t_f^-$, we have $g(0)=\dot g(0)=(L_Qg)(0)=0$ and $(\partial_\tau L_Qg)(0)=1$, so only the
$D_2$ terms with $\partial_\tau L_Qg$ survive. The bulk-boundary entries reduce to boundary ones,
\begin{align}
  \partial_tC_{qq}(t,t_f)&\to\Sigma_{12},\qquad \partial_t\partial_{t_f}C_{qq}(t,t_f)\to\Sigma_{22}, \label{eq: bulk_boundary}
\end{align}
so the braces become $D_2\big(\Sigma_{11}\Sigma_{22}-\Sigma_{12}^2\big)=D_2\det\Sigma$, and
\begin{align}
  \frac{d}{dt}\big\langle H_c\big\rangle^{\rm bridge}_{\rm conn}\Big|_{t\to t_f^-}
  &=\tfrac12\big(\alpha^2D_0+D_2\big)-D_2=\tfrac12\big(\alpha^2D_0-D_2\big). \label{eq: bridge_rate_tf}
\end{align}
Near the final time, the conditioning on the classical system thus reduces the net rate of heating, flipping the sign of the $D_2$ contribution from $+D_2$ to $-D_2$. The energies and heating rates are shown in Fig.~\ref{fig: energy}.

The full bridge energy is given by
\begin{align}
  \big\langle H\big\rangle^{\rm bridge}
  &=\underbrace{\langle H_c\big\rangle^{\rm bridge}_{\rm conn}+\tfrac18\,\Delta_2^{\rm conn}}_{\text{noise, independent of initial state}}
  +\underbrace{H_c\big[X_{\rm cl}\big]+\tfrac18\,\Delta_2^{\rm cl}}_{\text{initial state and record}}, \label{eq: H_bridge_full}
\end{align}
with
\begin{align}
  \Delta_2^{\rm conn}(t)&=\frac{\alpha^2}{\det\Sigma}\Big[\Sigma_{22}\big(\dot g^{2}+\omega_Q^2g^2\big)-2\Sigma_{12}\big(\dot g\ddot g+\omega_Q^2g\dot
  g\big)+\Sigma_{11}\big(\ddot g^{2}+\omega_Q^2\dot g^{2}\big)\Big]\;\ge0, \label{eq: Delta2_conn}\\
  H_c\big[X_{\rm cl}\big]+\tfrac18\,\Delta_2^{\rm cl}&=\tfrac12\big(\dot Q^{2}_{c,\rm cl}+\dot q^2_{\rm cl}\big)+\tfrac12\big(\omega_Q^2Q^{2}_{c,\rm cl}+\omega_c^2q^2_{\rm cl}\big)\notag\\
  &+\alpha\,Q_{c,\rm cl}\, q_{\rm cl}+\tfrac18\Big(\dot Q_{q,\rm cl}^{\,2}+\omega_Q^2\,Q_{q,\rm cl}^2\Big), \label{eq: H_cl}
\end{align}
the full heating rate consequently being given by
\begin{align}
  \frac{d}{dt}\big\langle H\big\rangle^{\rm bridge}
  &=\tfrac12\big(\alpha^2D_0+D_2\big)\notag\\
  &-\frac{1}{\det\Sigma}\Big\{
  \Sigma_{22}\Big[-\alpha^3D_0\,g\,\partial_tC_{Q_cq}+D_2\,(L_Qg)\,\partial_tC_{qq}\Big]\notag\\
  &-\Sigma_{12}\Big[-\alpha^3D_0\big(\dot g\,\partial_tC_{Q_cq}+g\,\partial_t\partial_{t_f}C_{Q_cq}\big)
  +D_2\big((\partial_\tau L_Qg)\,\partial_tC_{qq}+(L_Qg)\,\partial_t\partial_{t_f}C_{qq}\big)\Big]\notag\\
  &+\Sigma_{11}\Big[-\alpha^3D_0\,\dot g\,\partial_t\partial_{t_f}C_{Q_cq}+D_2\,(\partial_\tau L_Qg)\,\partial_t\partial_{t_f}C_{qq}\Big]\Big\}\notag\\
  &-\frac{\alpha^2}{4\det\Sigma}\Big[\Sigma_{22}\,\dot g\,(L_Qg)
  -\Sigma_{12}\big(\ddot g\,(L_Qg)+\dot g\,(\partial_\tau L_Qg)\big)
  +\Sigma_{11}\,\ddot g\,(\partial_\tau L_Qg)\Big]\notag\\
  &+\dot Q_{c,\rm cl}\big(\ddot Q_{c,\rm cl}+\omega_Q^2Q_{c,\rm cl}+\alpha\,q_{\rm cl}\big)
  +\dot q_{\rm cl}\big(\ddot q_{\rm cl}+\omega_c^2q_{\rm cl}+\alpha\,Q_{c,\rm cl}\big)\notag\\
  &+\tfrac14\,\dot Q_{q,\rm cl}\big(\ddot Q_{q,\rm cl}+\omega_Q^2Q_{q,\rm cl}\big). \label{eq: H_bridge_rate_full}
\end{align}

We thus see that the free case is equivalent to that of a pair of coupled classical harmonic oscillators; it is in the bridge that we notice quantum effects, even for a Gaussian system. The reason is simple: for a classical system, the response variable is a mathematical artifice, the physical variable being $q$, and thus it does not enter the definition of $H$. This is not the case for the quantum system, where both $Q_c$ and $Q_q$ are physical.

\section{The CQ oscillator as a measurement and feedback system}
\label{sec: measurement}

In this section, we explicitly formulate the CQ oscillator as a continuous weak-measurement and feedback system, with the classical degrees of freedom serving as a noisy meter. We use the Kalman--Bucy filter, the optimal continuous-time state estimator for linear dynamical systems with Gaussian process and observation noise, to estimate the state of the quantum oscillator. We show that, for a Gaussian initial state, the conditioned purity relaxes monotonically to $\sqrt{\eta}$, with $0<\eta\leq 1$ fixed by the diffusion-decoherence trade-off and $\eta=1$ indicating saturation. In addition, we show that the \emph{unconditional} purity goes to zero, non-monotonically in general, at a rate determined by the magnitude of the squared coupling constant $\alpha^2$ relative to the product of the squared frequencies, $\omega_Q^2\omega_c^2$.

For a linear system observed through a noisy linear channel,
\begin{align}
  dx&=\big(Ax+Bu\big)dt+dw,\qquad
  dy=Cx\,dt+dv, \label{eq: KB_system}\\
  \mathbb E\big[dw\,dw^{\mathsf T}\big]&=Q_S\,dt,\qquad
  \mathbb E\big[dv\,dv^{\mathsf T}\big]=R\,dt, \label{eq: KB_noise}
\end{align}
with $dw$ and $dv$ independent and $R\succ0$, the estimate $\hat{x}$ and the error covariance matrix $P$ obey
\begin{align}
  d\hat x&=\big(A\hat x+Bu\big)dt+K\big(dy-C\hat x\,dt\big),
  \qquad
  K=PC^{\mathsf T}R^{-1}, \label{eq: KB_filter}\\
  \dot P&=AP+PA^{\mathsf T}+Q_S-PC^{\mathsf T}R^{-1}CP, \label{eq: KB_Riccati}
\end{align}
where $K$ is the Kalman gain, the optimal weighting factor obtained by minimising the post-measurement error covariance matrix: the larger $K$ is, the more the meter can be relied upon for state estimation. Note that $P$ is deterministic and that, as $R\rightarrow0$, $P\rightarrow0$ such that $K$ remains bounded.

Let us now apply this formalism to the oscillator. With $P_{\rm cl}=\dot Q_c$ and $p=\dot q$ the conjugate momenta, we have
\begin{align}
  dQ_c&=P_{\rm cl}\,dt, & dP_{\rm cl}&=-\big(\omega_Q^2Q_c+\alpha q\big)dt+\alpha\sqrt{D_0}\,dW_1, \label{eq: SDE_Q}\\
  dq&=p\,dt, & dp&=-\big(\omega_c^2q+\alpha Q_c\big)dt+\sqrt{D_2}\,dW_2, \label{eq: SDE_q}
\end{align}
with $W_1,W_2$ independent standard Brownian motions, so that $\langle\xi_Q\xi_Q\rangle=\alpha^2D_0\delta$
and $\langle\xi_q\xi_q\rangle=D_2\delta$ as before. Then
\begin{align}
dx&=\big(A_Sx+Bu\big)dt+dw, \label{eq: KB_osc_state}\\
dy&\equiv-\frac{1}{\alpha}\Big[dp+\omega_c^2q\,dt\Big]
  =Cx\,dt+dv, \label{eq: KB_osc_obs}
\end{align}
where
\begin{align}
   x &=(Q_c,\ P_{\rm cl})^{\mathsf T}, & A_S&=\begin{pmatrix}0&1\\-\omega_Q^2&0\end{pmatrix}, & Bu&=\big(0,\,-\alpha q(t)\big)^{\mathsf T}, \label{eq: KB_osc_defs_1}\\
   C&= (1,\ 0), & Q_S&=\mathrm{diag}(0,\alpha^2D_0), & R&=\frac{D_2}{\alpha^2}. \label{eq: KB_osc_defs_2}
\end{align}
The estimation error $e=x-\hat x$ then obeys
\begin{align}
de&=(A_S-KC)e\,dt+dw-K\sqrt{R}\,dB, \label{eq: KB_error}
\end{align}
with $B$ a standard Brownian motion.
Substituting $R^{-1}=\alpha^2/D_2$ gives
\begin{align}
  \dot P&=A_SP+PA_S^{\mathsf T}+Q_S-\frac{\alpha^2}{D_2}\,PC^{\mathsf T}CP,
  \qquad
  K=PC^{\mathsf T}R^{-1}=\frac{\alpha^2}{D_2}\begin{pmatrix}P_{11}\\ P_{21}\end{pmatrix}, \label{eq:Riccati}\\
  d\hat x&=\Big(A_S\hat x+\big(0,\,-\alpha q(t)\big)^{\mathsf T}\Big)dt+K\big(dy-\hat x_1\,dt\big). \label{eq: KB_osc_filter}
\end{align}

For a Gaussian state, one can use $P$ to find the purity of the state conditioned on the classical trajectory. With the conditional
mean $\hat x$ and covariance $P$ on phase space $x=(Q_c,P_{\rm cl})$, the Wigner function is the
normalised Gaussian
\begin{align}
  W(x)&=\frac{1}{2\pi\sqrt{\det P}}\exp\!\Big[-\tfrac12(x-\hat x)^{\mathsf T}P^{-1}(x-\hat x)\Big], \label{eq: Wigner}
\end{align}
and the purity
is given by $\operatorname{Tr}\varrho^2=2\pi\int W^2\,d^2x$. Then
\begin{align}
  \operatorname{Tr}\varrho^2&=2\pi\cdot\frac{\pi\sqrt{\det P}}{4\pi^2\det P}
  =\frac{1}{2\sqrt{\det P}} . \label{eq: purity_Gaussian}
\end{align}
Note that the purity depends only on $P$ and is therefore the same on every trajectory. In addition, for
$\det P=1/4$ the state is pure, while for $\det P>1/4$ it is mixed; the Robertson--Schr\"odinger uncertainty relation forbids
anything below $1/4$. Finally, for non-Gaussian states the covariance matrix only gives the bound $\operatorname{Tr}\varrho^2\ge1 /(2\sqrt{\det P})$.

\begin{figure}[t]
  \centering
  \includegraphics[width=\linewidth]{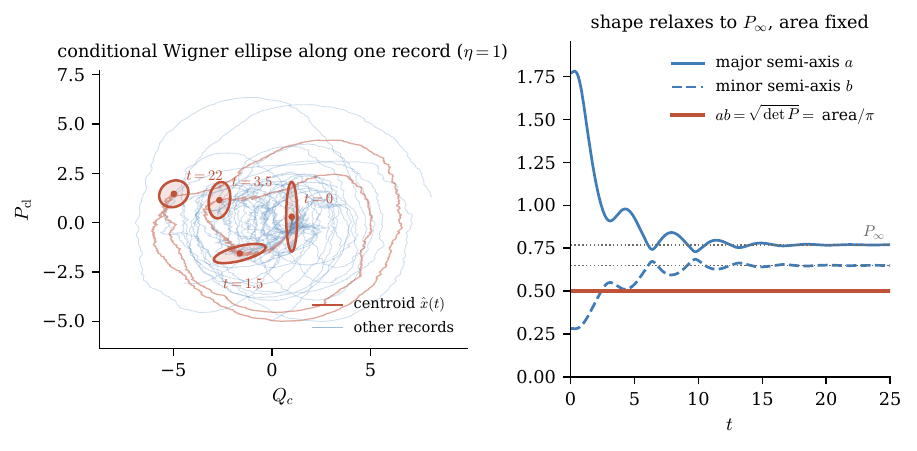}
  \caption{Conditional state of the quantum oscillator monitored by the classical one, Eqs.~\eqref{eq:Riccati} and~\eqref{eq: KB_osc_filter}, at saturation of the trade-off ($\eta=1$, $D_0D_2=1/4$), with the other parameters as in Fig.~\ref{fig: langevin}. The initial state is the squeezed pure state $P(0)=\operatorname{diag}(0.08,3.125)$, with initial estimate $\hat x(0)=(1,0.3)$ and meter data $(q,\dot q)(0)=(0.5,-0.2)$; system and filter are integrated by Euler--Maruyama with $\Delta t=10^{-3}$. Left: $1\sigma$ contours of the conditional Wigner function~\eqref{eq: Wigner} at the times shown, centred on the estimate $\hat x(t)$ along one record (red); other records in blue. Right: the semi-axes $a\geq b$ of the contour relax to those of the steady state $P_\infty$ (dotted), while $ab=\sqrt{\det P}$, the area divided by $\pi$, stays at $1/2$: the state remains pure.}
  \label{fig: conditional_state}
\end{figure}

\begin{figure}[t]
  \centering
  \includegraphics[width=0.55\linewidth]{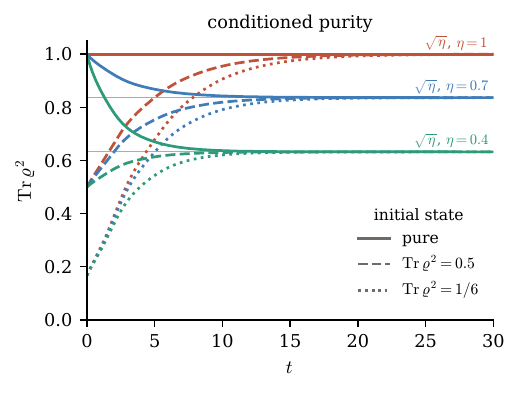}
  \caption{Relaxation of the conditioned purity $\operatorname{Tr}\varrho^2=1/(2\sqrt{\det P})$, from the Riccati equation~\eqref{eq:Riccati} with $D_0=1/(4\eta D_2)$, for $\eta=1$, $0.7$ and $0.4$ (red, blue, green), with $\omega_Q=0.9$, $\omega_c=1.3$, $\alpha=0.7$ and $D_2=1$. Initial states: the ground state of the free quantum oscillator, $P(0)=\operatorname{diag}\big(1/(2\omega_Q),\omega_Q/2\big)$ (solid), and the mixed states $P(0)=\mathbb 1$ (dashed, $\operatorname{Tr}\varrho^2=1/2$) and $P(0)=3\,\mathbb 1$ (dotted, $\operatorname{Tr}\varrho^2=1/6$). Every curve relaxes monotonically to $\sqrt\eta$, Eq.~\eqref{eq: detP_flow}; the purity is the same on every record.}
  \label{fig: purity}
\end{figure}

We can now use Eq.~\eqref{eq:Riccati} to examine the evolution of the purity of a weakly monitored system with Gaussian initial conditions:
\begin{align}
\frac{d}{dt}\det P&=K_1\big(D_0D_2-\det P\big),\qquad K_1=\frac{\alpha^2P_{11}}{D_2}, \label{eq: detP_flow}
\end{align}
where $K_1$ is the position Kalman gain. Then $\det P=D_0D_2$ is an invariant manifold, globally attracting for
$P_{11}>0$. With $D_0D_2=1/(4\eta)$, saturating the decoherence-diffusion trade-off ($\eta=1$)
makes the pure-state manifold invariant: a state that starts pure and Gaussian stays so for all
time, while an impure initial state monotonically purifies. Off saturation, the purity of the conditioned
state relaxes monotonically to $\sqrt{\eta}$ at the same rate, $K_1$. This is illustrated in Figs.~\ref{fig: conditional_state} and~\ref{fig: purity}.

The unconditional purity, on the other hand, declines, but the decline is monotonic only under certain conditions, and the form of $\operatorname{Tr}\varrho^2 (t)$ depends on the magnitude of $\alpha^2$ relative to $\omega_Q^2\omega_c^2$. To compute the purity explicitly as a function of $t$, we solve the Lyapunov equation
\begin{align}
    \dot\Sigma_L&=F\Sigma_L+\Sigma_L F^{\mathsf T}+\mathbf D, \qquad F=\begin{pmatrix}0&1&0&0\\-\omega_Q^2&0&-\alpha&0\\0&0&0&1\\-\alpha&0&-\omega_c^2&0\end{pmatrix},
  \qquad \mathbf D=\operatorname{diag}(0,\alpha^2D_0,0,D_2), \label{eq: Lyapunov}
\end{align}
where $\Sigma_L$ is the full $4\times4$ covariance matrix and $F$ is the drift matrix in the $x=(Q_c,P_{\rm cl},q,p)^{\mathsf T}$ basis, and extract the determinant of the $(Q_c,P_{\rm cl})$ subsystem to obtain the unconditional purity. The Lyapunov equation has the solution
\begin{align}
\Sigma_L(t)&=e^{Ft}\Sigma_{L,0}e^{F^{\mathsf T}t}+\int_0^te^{Fs}\mathbf De^{F^{\mathsf T}s}ds. \label{eq: Lyapunov_sol}
\end{align}
Matters may be considerably simplified by using spectral projectors: for
$\alpha^2\neq\omega_Q^2\omega_c^2$, $F$ is
diagonalisable, with right and left eigenvectors $Fv_\lambda=\lambda v_\lambda$ and
$w_\lambda^{\mathsf T}F=\lambda w_\lambda^{\mathsf T}$, $w_\lambda^{\mathsf T}v_\mu=\delta_{\lambda\mu}$, and the
projectors
$\Pi_\lambda=v_\lambda w_\lambda^{\mathsf T}$ satisfy $\Pi_\lambda\Pi_\mu=\delta_{\lambda\mu}\Pi_\lambda$,
$\sum_\lambda\Pi_\lambda=I$, and $e^{Fs}=\sum_\lambda e^{\lambda s}\Pi_\lambda$. Hence
\begin{align}
  \int_0^te^{Fs}\mathbf De^{F^{\mathsf T}s}ds&=\sum_{\lambda,\mu}\Pi_\lambda \mathbf D\Pi_\mu^{\mathsf T}\int_0^te^{(\lambda+\mu)s}ds, \label{eq: noise_integral}
\end{align}
and therefore
\begin{align}
  \Sigma_L(t)&=\sum_{\lambda,\mu}e^{(\lambda+\mu)t}\,\Pi_\lambda\Sigma_{L,0}\Pi_\mu^{\mathsf T}
  +\sum_{\lambda+\mu\neq0}\frac{e^{(\lambda+\mu)t}-1}{\lambda+\mu}\,\Pi_\lambda \mathbf D\Pi_\mu^{\mathsf T}
  +t\sum_\lambda\Pi_\lambda \mathbf D\,\Pi_{-\lambda}^{\mathsf T}. \label{eq: Sigma_spectral}
\end{align}
We can now compute the late-time asymptotics in the three regimes. Since the computations for the subcritical case are straightforward and indicative of the general case, we include them here, while only quoting the results for the critical and supercritical cases (the computations for those are more tedious than illuminating, and are performed in full in Appendix~\ref{app: purity}).

\emph{Subcritical}, $\alpha^2<\omega_Q^2\omega_c^2$, spectrum $\{\pm i\omega_+,\pm i\omega_-\}$: every
eigenvalue is imaginary, so
\begin{align}
  t\sum_\lambda\Pi_\lambda \mathbf D\Pi_{-\lambda}^{\mathsf T}&=t\,\bar\Sigma_L,\qquad
  \bar\Sigma_L=\sum_\lambda\Pi_\lambda \mathbf D\Pi_\lambda^\dagger=\sum_\lambda c_\lambda v_\lambda v_\lambda^\dagger\succeq0,
  \qquad c_\lambda=w_\lambda^\dagger \mathbf Dw_\lambda . \label{eq: Sigma_bar}
\end{align}
We can therefore write the covariance matrix as $\Sigma_L(t)=B(t)+t\bar{\Sigma_L}$, where $B(t)$ is a bounded matrix. To extract the submatrix $\Sigma_{L,S}$ of the quantum system, we use
$\Sigma_{L,S}=E\Sigma_L E^{\mathsf T}$ with
$E=(I_2\;0)$; then $\Sigma_L=t\bar\Sigma_L+B$ gives $\Sigma_{L,S}=t\bar\Sigma_{L,S}+B_S$, with $B_S$
bounded. Since our objective is $\det\Sigma_{L,S}$, and since for $2\times2$ matrices
$\det(tA+B)=t^2\det A+t\operatorname{tr}[\operatorname{adj}(A)B]+\det B$ holds exactly, to find the leading-order term in $t$ it is sufficient to compute $E(\sum_\lambda  c_\lambda v_\lambda v_\lambda^\dagger) E^{\mathsf T}$. The eigenvectors of $F$ are given by
\begin{align}
  v_\lambda&=(1,\lambda,r_\lambda,\lambda r_\lambda)^{\mathsf T},\qquad
  w_\lambda=\frac{(\lambda,1,\lambda r_\lambda,r_\lambda)^{\mathsf T}}{2\lambda(1+r_\lambda^2)},\qquad
  r_\lambda=-\frac{\lambda^2+\omega_Q^2}{\alpha}, \label{eq: eigvecs}
\end{align}
so that $Ev_{\lambda}=(v_\lambda)_S=(1,\lambda)$, with
$c_\lambda=w_\lambda^\dagger \mathbf Dw_\lambda=\frac{\alpha^2D_0+D_2r_\lambda^2}{4|\lambda|^2(1+r_\lambda^2)^2}>0$. Then
\begin{align}
  \bar\Sigma_{L,S}&=\sum_\lambda c_\lambda\begin{pmatrix}1&\bar\lambda\\\lambda&|\lambda|^2\end{pmatrix}, \label{eq: Sigma_bar_S}
\end{align}
whose off-diagonal entries cancel, leaving
\begin{align}
  \bar\sigma_{11}&=\sum_\pm\frac{\alpha^2D_0+D_2r_\pm^2}{2\omega_\pm^2(1+r_\pm^2)^2},\qquad
  \bar\sigma_{22}=\sum_\pm\frac{\alpha^2D_0+D_2r_\pm^2}{2(1+r_\pm^2)^2},\qquad
  r_\pm=\frac{\omega_\pm^2-\omega_Q^2}{\alpha}. \label{eq: sigma_bar_entries}
\end{align}
Therefore,
\begin{align}
\det\Sigma_{L,S}&=t^2\bar\sigma_{11}\bar\sigma_{22}+O(t),\qquad
  \operatorname{Tr}\varrho_S^2\simeq \frac{1}{2t\sqrt{\bar\sigma_{11}\bar\sigma_{22}}}. \label{eq: detSigma_sub}
\end{align}

\emph{Critical}, $\alpha^2=\omega_Q^2\omega_c^2$, spectrum $\{0,\pm i\omega_+\}$:
\begin{align}
\det\Sigma_{L,S}&\simeq\frac{(\omega_Q^2\omega_c^2)^2}{(\omega_Q^2+\omega_c^2)^4}\,
  (\omega_c^4D_0+D_2)\,\frac{(\omega_c^4D_0+D_2)+2(\omega_Q^4D_0+D_2)}{12}\;t^4,
  \qquad
  \operatorname{Tr}\varrho_S^2=\frac{1}{2\sqrt{\det\Sigma_{L,S}}}\propto t^{-2}. \label{eq: detSigma_crit}
\end{align}

\emph{Supercritical}, $\alpha^2>\omega_Q^2\omega_c^2$, spectrum $\{\pm i\omega_+,\pm\kappa\}$:
with
\begin{align}
  \gamma&=\frac{1}{(\omega_+^2+\kappa^2)^2}\left[
  \frac{(\omega_c^2+\kappa^2)^2\big(\kappa^2P_{0,11}+2\kappa P_{0,12}+P_{0,22}\big)}{4\kappa^2}
  +\frac{\alpha^2\big[D_0(\omega_c^2+\kappa^2)^2+D_2\big]}{8\kappa^3}\right], \label{eq: gamma}
\end{align}
we find
\begin{align}
\det\Sigma_{L,S}&=\gamma\,s_\perp\,t\,e^{2\kappa t}\big(1+O(t^{-1})\big),\qquad
  s_\perp=\frac{\alpha^2\big[D_0(\omega_Q^2+\kappa^2)^2+D_2\big]}{2\omega_+^2(\omega_+^2+\kappa^2)},
\qquad\operatorname{Tr}\varrho_S^2\simeq\frac{1}{2\sqrt{\gamma s_\perp}}\,\frac{e^{-\kappa t}}{\sqrt t}. \label{eq: detSigma_super}
\end{align}

\emph{When the unconditional purity declines monotonically.} Unconditionally, the system is not
Markovian on its own, so we work with the full state $x=(Q_c,P_{\rm cl},q,p)$, whose covariance obeys
the Lyapunov equation~\eqref{eq: Lyapunov}. The
unconditional purity is $1/(2\sqrt{\det\Sigma_{L,S}})$, with $\Sigma_{L,S}$ the $(Q_c,P_{\rm cl})$ block.
Differentiating once more gives $\ddot\Sigma_L=F\dot\Sigma_L+\dot\Sigma_L F^{\mathsf T}$, hence
\begin{align}
  \dot\Sigma_L(t)&=e^{Ft}\,\dot\Sigma_L(0)\,e^{F^{\mathsf T}t}. \label{eq: Sigma_dot}
\end{align}
If $\dot\Sigma_L(0)\succeq0$, then $\Sigma_L(t)$ is non-decreasing in the Loewner order, and so is its
block $\Sigma_{L,S}(t)$. Since $B\succeq A\succ0$ implies $\det B\ge\det A$, $\det\Sigma_{L,S}$ is
non-decreasing and the purity is non-increasing. With the meter's Cauchy data exact,
$\Sigma_L(0)=\operatorname{diag}(P_0,0)$, the $q$ row of $\dot\Sigma_L(0)$ vanishes and the $p$ row is
$(-\alpha P_{11},-\alpha P_{12},0,D_2)$. A Schur complement on $p$ gives
\begin{align}
  \dot\Sigma_L(0)\succeq0
  &\iff
  A_SP_0+P_0A_S^{\mathsf T}+Q_S-\frac{\alpha^2}{D_2}P_0C^{\mathsf T}CP_0\succeq0 , \label{eq: monotone_condition}
\end{align}
which is the Riccati velocity at $P_0$. \emph{The unconditional purity declines monotonically whenever
the conditional covariance would initially grow in the Loewner order.} This holds for $P_0=P_\infty$ ($\dot P=0$), which is pure at $\eta=1$.


\section{Discussion}
\label{sec: discussion}

In this paper, we have constructed the CQ path integral for a quantum oscillator coupled to a classical one and obtained the semi-classical action, the expectation and weak values, and the associated two point functions. Using these, we were able to find exactly the expectation value of the energy, $H$ and the rate of heating. In addition, we re-cast the system as a measurement-feedback system and obtained analytical expressions for the purity of the system, both when conditioned on a classical trajectory, and when unconditioned. The expression for the unconditional case was obtained in three regimes: the sub-critical, critical, and super-critical regimes ($\alpha^2<\omega_Q^2\omega_c^2,\alpha^2=<\omega_Q^2\omega_c^2,\alpha^2\geq <\omega_Q^2\omega_c^2$) which correspond to weak, critical, and strong coupling between the classical and quantum systems.

We find that the unconditioned system has no stable state, and the energy grows linearly with $t$, whereas the growth in sub-linear in the bridge case. Nevertheless, it is clear that the addition of friction in either of the two systems would be sufficient to arrest this growth. It would therefore be a simple extension of this work to find the necessary conditions on friction for equilibrium steady states, and therefore the saturation energy and purity.

We would therefore like to better understand, for example, the exchange
of energy between the classical and quantum systems, as well as situations in which one of the oscillators is kept in a thermal state, as in the case which gives rise to the ultraviolet catastrophe. Another natural extension is to the anharmonic oscillator, where the Gaussian treatment is no longer sufficient. Finally, given that a motivation is to use such examples to understand gravity, modelling such systems in higher dimensions and in curved spaces is a natural extension. There, the Ito vs. Stratanovich prescriptions and ordering ambiguities may play a much more important role than in this case. Regardless, we hope that this simple model can be used to gain a fuller understanding of classical-quantum dynamics, both in the case of an effective theory~\cite{layton2023classical} and in the case of gravity, where it has been proposed as a candidate for a fundamental theory~\cite{oppenheim2018post,oppenheim2023covariant}.

%
\appendix

\section{Calculational details}
\label{app: details}
\subsection{Solving the equations of motion}
\label{app: EOM}

\emph{Step 1: the equation for $P$.}
The second equation of motion~\eqref{eq: eom_2} gives
$\ddot q+\omega_c^2q+\alpha Q_c=-\frac{D_2}{\alpha}\big(\ddot P+\omega_Q^2P\big)$.
Inserting this into the third,
\begin{align}
  \big(\partial_t^2+\omega_c^2\big)\big(\partial_t^2+\omega_Q^2\big)P-\alpha^2P&=0
  \quad\Longleftrightarrow\quad
  \big(\partial_t^2+\omega_+^2\big)\big(\partial_t^2+\omega_-^2\big)P=0 , \label{eq: P_equation}
\end{align}
Hence
\begin{align}
  P(t)&=c_1\cos\omega_+t+c_2\sin\omega_+t+c_3\cos\omega_-t+c_4\sin\omega_-t . \label{eq: P_sol_app}
\end{align}

\emph{Step 2: $Q_c$ and $q$ in terms of $P$.}
$Q_c$ and $q$ obey
\begin{align}
  \ddot Q_c+\omega_Q^2Q_c+\alpha q&=D_0\alpha^2P,\qquad
  \alpha Q_c+\ddot q+\omega_c^2q=-\frac{D_2}{\alpha}\big(\ddot P+\omega_Q^2P\big). \label{eq: Qc_q_from_P}
\end{align}
Let $Y$ be any solution of $(\partial_t^2+\omega_+^2)(\partial_t^2+\omega_-^2)Y=P$. Then
\begin{align}
  Q_c&=\Big[\alpha^2D_0\big(\partial_t^2+\omega_c^2\big)+D_2\big(\partial_t^2+\omega_Q^2\big)\Big]Y+(\text{homogeneous}), \label{eq: Qc_Y}\\
  q&=-\frac{1}{\alpha}\Big[D_2\big(\partial_t^2+\omega_Q^2\big)^2+\alpha^4D_0\Big]Y+(\text{homogeneous}). \label{eq: q_Y}
\end{align}
Substituting back, the left-hand sides become
$\alpha^2D_0\big[(\partial_t^2+\omega_Q^2)(\partial_t^2+\omega_c^2)-\alpha^2\big]Y=D_0\alpha^2P$ and
$-\frac{D_2}{\alpha}(\partial_t^2+\omega_Q^2)\big[(\partial_t^2+\omega_Q^2)(\partial_t^2+\omega_c^2)-\alpha^2\big]Y$,
so both equations hold.
$P$ is itself a solution of the homogeneous equation, so the forcing is resonant and $Y$ is secular:
\begin{align}
  Y&=\frac{t}{2\big(\omega_+^2-\omega_-^2\big)}\Big[\frac{-c_1\sin\omega_+t+c_2\cos\omega_+t}{\omega_+}
  -\frac{-c_3\sin\omega_-t+c_4\cos\omega_-t}{\omega_-}\Big]. \label{eq: Y_sol}
\end{align}
The homogeneous solutions are the normal modes, with $q=\frac{\omega_\pm^2-\omega_Q^2}{\alpha}Q_c$ in each mode.
Carrying out the derivatives gives
\begin{align}
  Q_c(t)&=\frac{1}{2(\omega_+^2-\omega_-^2)}\Big\{t\,\big[\alpha^2D_0(\omega_c^2-\omega_+^2)+D_2(\omega_Q^2-\omega_+^2)\big]\frac{-c_1\sin\omega_+t+c_2\cos\omega_+t}{\omega_+}\notag\\
  &-2(\alpha^2D_0+D_2)\big(c_1\cos\omega_+t+c_2\sin\omega_+t\big)\notag\\
  &-t\,\big[\alpha^2D_0(\omega_c^2-\omega_-^2)+D_2(\omega_Q^2-\omega_-^2)\big]\frac{-c_3\sin\omega_-t+c_4\cos\omega_-t}{\omega_-}\notag\\
  &+2(\alpha^2D_0+D_2)\big(c_3\cos\omega_-t+c_4\sin\omega_-t\big)\Big\}\notag\\
  &+c_5\cos\omega_+t+c_6\sin\omega_+t+c_7\cos\omega_-t+c_8\sin\omega_-t, \label{eq: Qc_sol_app}\\[6pt]
  q(t)&=-\frac{1}{2\alpha(\omega_+^2-\omega_-^2)}\Big\{t\,\big[D_2(\omega_Q^2-\omega_+^2)^2+\alpha^4D_0\big]\frac{-c_1\sin\omega_+t+c_2\cos\omega_+t}{\omega_+}\notag\\
  &-4D_2(\omega_Q^2-\omega_+^2)\big(c_1\cos\omega_+t+c_2\sin\omega_+t\big)\notag\\
  &-t\,\big[D_2(\omega_Q^2-\omega_-^2)^2+\alpha^4D_0\big]\frac{-c_3\sin\omega_-t+c_4\cos\omega_-t}{\omega_-}\notag\\
  &+4D_2(\omega_Q^2-\omega_-^2)\big(c_3\cos\omega_-t+c_4\sin\omega_-t\big)\Big\}\notag\\
  &+\frac{\omega_+^2-\omega_Q^2}{\alpha}\big(c_5\cos\omega_+t+c_6\sin\omega_+t\big)
  +\frac{\omega_-^2-\omega_Q^2}{\alpha}\big(c_7\cos\omega_-t+c_8\sin\omega_-t\big). \label{eq: q_sol_app}
\end{align}

\emph{Step 3: the eight conditions.}
Final data of $P$:
\begin{align}
  &c_1\cos\omega_+t_f+c_2\sin\omega_+t_f+c_3\cos\omega_-t_f+c_4\sin\omega_-t_f=0, \label{eq: cond_1}\\
  &\omega_+\big(-c_1\sin\omega_+t_f+c_2\cos\omega_+t_f\big)+\omega_-\big(-c_3\sin\omega_-t_f+c_4\cos\omega_-t_f\big)=0 . \label{eq: cond_2}
\end{align}
Initial data of $Q_c$. At $t=0$ the sines and the secular terms vanish:
\begin{align}
  &-\frac{\alpha^2D_0+D_2}{\omega_+^2-\omega_-^2}\,(c_1-c_3)+c_5+c_7=Q_i, \label{eq: cond_3}\\[4pt]
  &\frac{1}{2(\omega_+^2-\omega_-^2)}\Big\{\big[\alpha^2D_0(\omega_c^2-3\omega_+^2)+D_2(\omega_Q^2-3\omega_+^2)\big]\frac{c_2}{\omega_+}
  -\big[\alpha^2D_0(\omega_c^2-3\omega_-^2)+D_2(\omega_Q^2-3\omega_-^2)\big]\frac{c_4}{\omega_-}\Big\}\notag\\
  &\qquad+\omega_+c_6+\omega_-c_8=\dot Q_i . \label{eq: cond_4}
\end{align}
Initial data of $q$, here zero:
\begin{align}
  &\frac{2D_2}{\alpha(\omega_+^2-\omega_-^2)}\Big[(\omega_Q^2-\omega_+^2)\,c_1-(\omega_Q^2-\omega_-^2)\,c_3\Big]
  +\frac{\omega_+^2-\omega_Q^2}{\alpha}\,c_5+\frac{\omega_-^2-\omega_Q^2}{\alpha}\,c_7=0, \label{eq: cond_5}\\[4pt]
  &-\frac{1}{2\alpha(\omega_+^2-\omega_-^2)}\Big\{\big[D_2(\omega_Q^2-\omega_+^2)(\omega_Q^2-5\omega_+^2)+\alpha^4D_0\big]\frac{c_2}{\omega_+}
  -\big[D_2(\omega_Q^2-\omega_-^2)(\omega_Q^2-5\omega_-^2)+\alpha^4D_0\big]\frac{c_4}{\omega_-}\Big\}\notag\\
  &\qquad+\frac{(\omega_+^2-\omega_Q^2)\,\omega_+}{\alpha}\,c_6+\frac{(\omega_-^2-\omega_Q^2)\,\omega_-}{\alpha}\,c_8=0 . \label{eq: cond_6}
\end{align}
Final data of $q$, where the secular terms survive:
\begin{align}
  q(t_f)&=q_f, \label{eq: cond_7}\\
  \dot q(t_f)&=\dot q_f, \label{eq: cond_8}
\end{align}
with $q(t)$ given by Eq.~\eqref{eq: q_sol_app}, and
\begin{align}
  \dot q(t)&=-\frac{1}{2\alpha(\omega_+^2-\omega_-^2)}\Big\{\big[D_2(\omega_Q^2-\omega_+^2)^2+\alpha^4D_0\big]\Big[\frac{-c_1\sin\omega_+t+c_2\cos\omega_+t}
  {\omega_+}-t\big(c_1\cos\omega_+t+c_2\sin\omega_+t\big)\Big]\notag\\
  &-4D_2(\omega_Q^2-\omega_+^2)\,\omega_+\big(-c_1\sin\omega_+t+c_2\cos\omega_+t\big)\notag\\
  &-\big[D_2(\omega_Q^2-\omega_-^2)^2+\alpha^4D_0\big]\Big[\frac{-c_3\sin\omega_-t+c_4\cos\omega_-t}{\omega_-}-t\big(c_3\cos\omega_-t+c_4\sin\omega_-t\big)\Big]\notag\\
  &+4D_2(\omega_Q^2-\omega_-^2)\,\omega_-\big(-c_3\sin\omega_-t+c_4\cos\omega_-t\big)\Big\}\notag\\
  &+\frac{(\omega_+^2-\omega_Q^2)\,\omega_+}{\alpha}\big(-c_5\sin\omega_+t+c_6\cos\omega_+t\big)
  +\frac{(\omega_-^2-\omega_Q^2)\,\omega_-}{\alpha}\big(-c_7\sin\omega_-t+c_8\cos\omega_-t\big). \label{eq: qdot_sol}
\end{align}
These are eight linear equations for $c_1,\dots,c_8$.

\subsection{Two-point functions}
\label{app: 2pf}
\emph{The pinned partition function.}
Let $\Phi=(Q_c,Q_q,q)$, let $\langle J,\Phi\rangle=\int_0^{t_f} dt\,\sum_XJ_X(t)X(t)$, and let
the action be $-\tfrac12\int\Phi^{\mathsf T}\mathcal M\Phi$. Paths have their initial data
prescribed, and the bridge pins the final meter data:
\begin{align}
  Z[J;Y]&=\int_{\rm free}\!\mathcal D\Phi\;e^{-\frac12\int\Phi^{\mathsf T}\mathcal M\Phi+\langle J,\Phi\rangle}\;
  \delta\big(q(t_f)-Y_1\big)\,\delta\big(\dot q(t_f)-Y_2\big). \label{eq: Z_pinned}
\end{align}
Here ``free'' denotes the free path space, in which $q(t_f)$ and $\dot q(t_f)$ are unconstrained. It is
normalised so that
\begin{align}
  Z_{\rm free}[J]&=\int_{\rm free}\!\mathcal D\Phi\;e^{-\frac12\int\Phi^{\mathsf T}\mathcal M\Phi+\langle J,\Phi\rangle}
  =e^{\frac12\langle J,CJ\rangle+\langle J,\Phi_{\rm det}\rangle},\qquad Z_{\rm free}[0]=1 . \label{eq: Z_free}
\end{align}
$C$ is the Green's function of $\mathcal M$ with the free (natural) conditions at $t_f$, and
$\Phi_{\rm det}$ is the noiseless solution.

\emph{Construction A: free process plus delta function (conditioning).}
Writing each $\delta$ as a Fourier integral,
\begin{align}
  Z[J;Y]&=\int\!\frac{d\beta_1\,d\beta_2}{(2\pi)^2}\;e^{-i\beta_1Y_1-i\beta_2Y_2}\;
  Z_{\rm free}\big[J+i\beta_1\delta_q(\cdot-t_f)+i\beta_2\delta_{\dot q}(\cdot-t_f)\big]. \label{eq: Z_Fourier}
\end{align}
The extra source couples to $q(t_f)$ and $\dot q(t_f)$. Inserting the free result and using the
definitions of $\Sigma$ and $\mathbf h$ (Eqs.~\eqref{eq: Sigma_h} and~\eqref{eq: h_def}),
\begin{align}
  \big\langle\delta_q(\cdot-t_f),\,C\,\delta_q(\cdot-t_f)\big\rangle&=\Sigma_{11},\ \ldots,\qquad
  \big\langle\delta_q(\cdot-t_f),\,CJ\big\rangle=\langle\mathbf h_1,J\rangle,\quad
  \big\langle\delta_{\dot q}(\cdot-t_f),\,CJ\big\rangle=\langle\mathbf h_2,J\rangle, \label{eq: source_overlaps}
\end{align}
where $\langle\mathbf h_a,J\rangle=\int dt\sum_X\mathbf h_X(t)_aJ_X(t)$. The exponent becomes
\begin{align}
  &\tfrac12\langle J,CJ\rangle+\langle J,\Phi_{\rm det}\rangle
  -\tfrac12\beta^{\mathsf T}\Sigma\beta
  +i\beta^{\mathsf T}\big(\langle\mathbf h,J\rangle-\delta Y\big),
  \qquad \delta Y=Y-\big(q_{\rm det}(t_f),\dot q_{\rm det}(t_f)\big). \label{eq: exponent_A}
\end{align}
The $\beta$-integral is Gaussian with $\Sigma\succ0$:
\begin{align}
  \boxed{\;
  Z[J;Y]=\frac{e^{-\frac12\delta Y^{\mathsf T}\Sigma^{-1}\delta Y}}{2\pi\sqrt{\det\Sigma}}\;
  \exp\Big\{\tfrac12\big\langle J,\big(C-\mathbf h\Sigma^{-1}\mathbf h^{\mathsf T}\big)J\big\rangle
  +\big\langle J,\Phi_{\rm det}+\mathbf h\Sigma^{-1}\delta Y\big\rangle\Big\}\;} \label{eq: Z_A}
\end{align}

\emph{Construction B: fluctuation field with bridge conditions (Green's function).}
On the pinned path space, write $\Phi=\Phi_{\rm cl}+\varphi$. Here $\Phi_{\rm cl}$ solves
$\mathcal M\Phi_{\rm cl}=0$ with all the data (initial data, $Y$, and $Q_q(t_f)=0$), and
$\varphi$ satisfies the homogeneous bridge conditions. The cross term is a boundary term that
vanishes under these conditions, so
\begin{align}
  -\tfrac12\int\Phi^{\mathsf T}\mathcal M\Phi&=S_{\rm cl}(Y)-\tfrac12\int\varphi^{\mathsf T}\mathcal M_{\rm pinned}\varphi . \label{eq: action_split}
\end{align}
Completing the square in $\varphi$ with $\mathcal G=\mathcal M_{\rm pinned}^{-1}$ (homogeneous bridge conditions),
\begin{align}
  \boxed{\;
  Z[J;Y]=Z_2[0]\;e^{S_{\rm cl}(Y)}\;
  \exp\Big\{\tfrac12\langle J,\mathcal G J\rangle+\langle J,\Phi_{\rm cl}\rangle\Big\}\;} \label{eq: Z_B}
\end{align}

\emph{Equivalence.}
Both boxed expressions are the same functional of $J$ for every $J$. Comparing $\ln Z$ order by order in $J$:
\begin{align}
  O(J^2):&\quad \mathcal G=C-\mathbf h\Sigma^{-1}\mathbf h^{\mathsf T}
  &&\text{(Green's function = conditional covariance)}, \label{eq: equiv_J2}\\
  O(J):&\quad \Phi_{\rm cl}=\Phi_{\rm det}+\mathbf h\Sigma^{-1}\delta Y
  &&\text{(boundary-value solution = conditional mean)}, \label{eq: equiv_J1}\\
  O(J^0):&\quad S_{\rm cl}=-\tfrac12\delta Y^{\mathsf T}\Sigma^{-1}\delta Y,\qquad
  Z_2[0]=\frac{1}{2\pi\sqrt{\det\Sigma}}
  &&\text{(on-shell action and fluctuation determinant)}. \label{eq: equiv_J0}
\end{align}
The last line is a Gel'fand--Yaglom-type statement: the ratio of the bridge and free
fluctuation determinants is the $2\times2$ determinant of $\Sigma$,
\begin{align}
  \frac{\det{}^{-1/2}\mathcal M_{\rm pinned}}{\det{}^{-1/2}\mathcal M}&=\frac{1}{2\pi\sqrt{\det\Sigma}} . \label{eq: Gelfand_Yaglom}
\end{align}
Integrating the boxed result over $Y$ returns $Z_{\rm free}[J]$, since the $\delta$-functions integrate
to one. This is the gluing identity, now exact for every $J$.

\emph{The vectors $\mathbf h_X$.} These are
\begin{align}
  \mathbf h_X(t)&=\Big(\big\langle X(t)\,q(t_f)\big\rangle,\ \big\langle X(t)\,\dot q(t_f)\big\rangle\Big). \label{eq: h_def}
\end{align}
The free solutions are
\begin{align}
  Q_c(t)&=\int_0^{t}\!ds\,\Big[\alpha\sqrt{D_0}\,(L_qg)(t-s)\,\xi_1(s)-\alpha\sqrt{D_2}\,g(t-s)\,\xi_2(s)\Big], \label{eq: Qc_free}\\
  q(t)&=\int_0^{t}\!ds\,\Big[-\alpha^2\sqrt{D_0}\,g(t-s)\,\xi_1(s)+\sqrt{D_2}\,(L_Qg)(t-s)\,\xi_2(s)\Big]. \label{eq: q_free}
\end{align}
We average over the two independent white noises. Because $t<t_f$, the integration stops at $t$.

Meter:
\begin{align}
  \big\langle q(t)\,q(t_f)\big\rangle&=\int_0^{t}\!ds\,\Big[\alpha^4D_0\,g(t-s)\,g(t_f-s)+D_2\,(L_Qg)(t-s)\,(L_Qg)(t_f-s)\Big], \label{eq: h_qq}\\
  \big\langle q(t)\,\dot q(t_f)\big\rangle&=\int_0^{t}\!ds\,\Big[\alpha^4D_0\,g(t-s)\,\dot g(t_f-s)+D_2\,(L_Qg)(t-s)\,(\partial_\tau L_Qg)(t_f-s)\Big]. \label{eq: h_qqdot}
\end{align}
Keldysh classical coordinate $Q_c$:
\begin{align}
  \big\langle Q_c(t)\,q(t_f)\big\rangle&=-\alpha\int_0^{t}\!ds\,\Big[\alpha^2D_0\,(L_qg)(t-s)\,g(t_f-s)+D_2\,g(t-s)\,(L_Qg)(t_f-s)\Big], \label{eq: h_Qcq}\\
  \big\langle Q_c(t)\,\dot q(t_f)\big\rangle&=-\alpha\int_0^{t}\!ds\,\Big[\alpha^2D_0\,(L_qg)(t-s)\,\dot g(t_f-s)+D_2\,g(t-s)\,(\partial_\tau L_Qg)(t_f-s)\Big]. \label{eq: h_Qcqdot}
\end{align}
Quantum (response) coordinate $Q_q$. Here the free correlator is a pure response function and
carries no noise:
\begin{align}
  \big\langle Q_q(t)\,q(t_f)\big\rangle&=i\alpha\,g(t_f-t),
  \qquad
  \big\langle Q_q(t)\,\dot q(t_f)\big\rangle=i\alpha\,\dot g(t_f-t). \label{eq: h_Qq}
\end{align}
In every case, the second component is the $t_f$-derivative of the first at fixed $t$,
\begin{align}
  \big\langle X(t)\,\dot q(t_f)\big\rangle&=\partial_{t_f}\big\langle X(t)\,q(t_f)\big\rangle , \label{eq: h_derivative}
\end{align}
since $t_f$ enters only through the boundary leg. At $t=t_f$, the meter entries reduce to the
boundary covariance:
\begin{align}
  \mathbf h_q(t_f)&=\Big(\langle q(t_f)q(t_f)\rangle,\ \langle q(t_f)\dot q(t_f)\rangle\Big),
  \qquad
  \partial_t\mathbf h_q(t)\big|_{t=t_f}=\Big(\langle \dot q(t_f)q(t_f)\rangle,\ \langle \dot q(t_f)\dot q(t_f)\rangle\Big). \label{eq: h_boundary}
\end{align}
The kernels are
\begin{align}
  g(\tau)&=\frac{\theta(\tau)}{\omega_+^2-\omega_-^2}\left[\frac{\sin\omega_-\tau}{\omega_-}-\frac{\sin\omega_+\tau}{\omega_+}\right], \label{eq: g_app}\\
  (L_qg)(\tau)&=\frac{\theta(\tau)}{\omega_+^2-\omega_-^2}\left[\big(\omega_+^2-\omega_c^2\big)\frac{\sin\omega_+\tau}{\omega_+}+\big(\omega_c^2-\omega_-^2\big)\frac{\sin\omega_-\tau}{\omega_-}\right], \label{eq: Lqg_app}\\
  (L_Qg)(\tau)&=\frac{\theta(\tau)}{\omega_+^2-\omega_-^2}\left[\big(\omega_+^2-\omega_Q^2\big)\frac{\sin\omega_+\tau}{\omega_+}+\big(\omega_Q^2-\omega_-^2\big)\frac{\sin\omega_-\tau}{\omega_-}\right]; \label{eq: LQg_app}.
\end{align}

\subsection{Purity}
\label{app: purity}
To compute the purity explicitly as a function of $t$, we solve the Lyapunov equation
\begin{align}
    \dot\Sigma_L&=F\Sigma_L+\Sigma_L F^{\mathsf T}+\mathbf D, \qquad F=\begin{pmatrix}0&1&0&0\\-\omega_Q^2&0&-\alpha&0\\0&0&0&1\\-\alpha&0&-\omega_c^2&0\end{pmatrix},
  \qquad \mathbf D=\operatorname{diag}(0,\alpha^2D_0,0,D_2), \label{eq: Lyapunov_app}
\end{align}
where $\Sigma_L$ is the full $4\times4$ covariance matrix and $F$ is the drift matrix in the $x=(Q_c,P_{\rm cl},q,p)^{\mathsf T}$ basis, and extract the determinant of the $(Q_c,P_{\rm cl})$ subsystem to obtain the unconditional purity.

\emph{Eigenvalues of $F$.} Before we proceed, let us write down the eigenvalues of $F$. The characteristic polynomial
$\det(F-\lambda)=(\lambda^2+\omega_Q^2)(\lambda^2+\omega_c^2)-\alpha^2$ is quadratic in $\lambda^2$, with
roots
\begin{align}
  \lambda^2&=-\frac12\Big[\omega_Q^2+\omega_c^2\pm\sqrt{(\omega_Q^2-\omega_c^2)^2+4\alpha^2}\,\Big]. \label{eq: lambda_sq}
\end{align}
Since
$(\omega_Q^2+\omega_c^2)^2-\big[(\omega_Q^2-\omega_c^2)^2+4\alpha^2\big]=4(\omega_Q^2\omega_c^2-\alpha^2)$,
the root with the upper sign is always negative, and the sign of the other is set by
$\omega_Q^2\omega_c^2-\alpha^2$.

\emph{Subcritical}, $\alpha^2<\omega_Q^2\omega_c^2$:
\begin{align}
  \lambda&=\pm\,i\sqrt{\tfrac12\Big[\omega_Q^2+\omega_c^2+\sqrt{(\omega_Q^2-\omega_c^2)^2+4\alpha^2}\,\Big]},
  \qquad
  \lambda=\pm\,i\sqrt{\tfrac12\Big[\omega_Q^2+\omega_c^2-\sqrt{(\omega_Q^2-\omega_c^2)^2+4\alpha^2}\,\Big]}. \label{eq: eig_sub}
\end{align}

\emph{Critical}, $\alpha^2=\omega_Q^2\omega_c^2$: here
$\sqrt{(\omega_Q^2-\omega_c^2)^2+4\alpha^2}=\omega_Q^2+\omega_c^2$, so
\begin{align}
  \lambda&=\pm\,i\sqrt{\omega_Q^2+\omega_c^2},\qquad \lambda=0\ \ (\text{double; one Jordan block}). \label{eq: eig_crit}
\end{align}

\emph{Supercritical}, $\alpha^2>\omega_Q^2\omega_c^2$:
\begin{align}
  \lambda&=\pm\,i\sqrt{\tfrac12\Big[\omega_Q^2+\omega_c^2+\sqrt{(\omega_Q^2-\omega_c^2)^2+4\alpha^2}\,\Big]},
  \qquad
  \lambda=\pm\sqrt{\tfrac12\Big[\sqrt{(\omega_Q^2-\omega_c^2)^2+4\alpha^2}-\omega_Q^2-\omega_c^2\Big]}. \label{eq: eig_super}
\end{align}
In every case the pair with the upper sign is purely imaginary. The soft pair is imaginary below
the threshold, zero at it, and real above it. Indeed, for the soft pair, $\lambda^2=\tfrac12\big[\sqrt{(\omega_Q^2-\omega_c^2)^2+4\alpha^2}-\omega_Q^2-\omega_c^2\big]$, where the inner
root is real since its radicand is a sum of squares. Squaring the non-negative quantities
$\sqrt{(\omega_Q^2-\omega_c^2)^2+4\alpha^2}$ and $\omega_Q^2+\omega_c^2$ shows that
$\lambda^2>0\iff\alpha^2>\omega_Q^2\omega_c^2$. Hence $\lambda=\pm\kappa$, with
\begin{align}
  \kappa&=\sqrt{\tfrac12\Big[\sqrt{(\omega_Q^2-\omega_c^2)^2+4\alpha^2}-\omega_Q^2-\omega_c^2\Big]}>0 . \label{eq: kappa}
\end{align}

The Lyapunov equation has the solution
\begin{align}
\Sigma_L(t)&=e^{Ft}\Sigma_{L,0}e^{F^{\mathsf T}t}+\int_0^te^{Fs}\mathbf De^{F^{\mathsf T}s}ds. \label{eq: Lyapunov_sol_app}
\end{align}
Matters may be considerably simplified by using spectral projectors: for
$\alpha^2\neq\omega_Q^2\omega_c^2$, $F$ is
diagonalisable, with right and left eigenvectors $Fv_\lambda=\lambda v_\lambda$ and
$w_\lambda^{\mathsf T}F=\lambda w_\lambda^{\mathsf T}$, $w_\lambda^{\mathsf T}v_\mu=\delta_{\lambda\mu}$, and the
projectors
$\Pi_\lambda=v_\lambda w_\lambda^{\mathsf T}$ satisfy $\Pi_\lambda\Pi_\mu=\delta_{\lambda\mu}\Pi_\lambda$,
$\sum_\lambda\Pi_\lambda=I$, and $e^{Fs}=\sum_\lambda e^{\lambda s}\Pi_\lambda$. Hence
\begin{align}
  \int_0^te^{Fs}\mathbf De^{F^{\mathsf T}s}ds&=\sum_{\lambda,\mu}\Pi_\lambda \mathbf D\Pi_\mu^{\mathsf T}\int_0^te^{(\lambda+\mu)s}ds, \label{eq: noise_integral_app}
\end{align}
and therefore
\begin{align}
  \Sigma_L(t)&=\sum_{\lambda,\mu}e^{(\lambda+\mu)t}\,\Pi_\lambda\Sigma_{L,0}\Pi_\mu^{\mathsf T}
  +\sum_{\lambda+\mu\neq0}\frac{e^{(\lambda+\mu)t}-1}{\lambda+\mu}\,\Pi_\lambda \mathbf D\Pi_\mu^{\mathsf T}
  +t\sum_\lambda\Pi_\lambda \mathbf D\,\Pi_{-\lambda}^{\mathsf T}. \label{eq: Sigma_spectral_app}
\end{align}
We now compute the late-time asymptotics in the three regimes.

\emph{Subcritical}, $\alpha^2<\omega_Q^2\omega_c^2$, spectrum $\{\pm i\omega_+,\pm i\omega_-\}$: every
eigenvalue is imaginary, so
\begin{align}
  t\sum_\lambda\Pi_\lambda \mathbf D\Pi_{-\lambda}^{\mathsf T}&=t\,\bar\Sigma_L,\qquad
  \bar\Sigma_L=\sum_\lambda\Pi_\lambda \mathbf D\Pi_\lambda^\dagger=\sum_\lambda c_\lambda v_\lambda v_\lambda^\dagger\succeq0,
  \qquad c_\lambda=w_\lambda^\dagger \mathbf Dw_\lambda . \label{eq: Sigma_bar_app}
\end{align}
We can therefore write the covariance matrix as $\Sigma_L(t)=B(t)+t\bar{\Sigma_L}$, where $B(t)$ is a bounded matrix. To extract the submatrix $\Sigma_{L,S}$ of the quantum system, we use
$\Sigma_{L,S}=E\Sigma_L E^{\mathsf T}$ with
$E=(I_2\;0)$; then $\Sigma_L=t\bar\Sigma_L+B$ gives $\Sigma_{L,S}=t\bar\Sigma_{L,S}+B_S$, with $B_S$
bounded. Since our objective is $\det\Sigma_{L,S}$, and since for $2\times2$ matrices
$\det(tA+B)=t^2\det A+t\operatorname{tr}[\operatorname{adj}(A)B]+\det B$ holds exactly, to find the leading-order term in $t$ it is sufficient to compute $E(\sum_\lambda  c_\lambda v_\lambda v_\lambda^\dagger) E^{\mathsf T}$. The eigenvectors of $F$ are given by
\begin{align}
  v_\lambda&=(1,\lambda,r_\lambda,\lambda r_\lambda)^{\mathsf T},\qquad
  w_\lambda=\frac{(\lambda,1,\lambda r_\lambda,r_\lambda)^{\mathsf T}}{2\lambda(1+r_\lambda^2)},\qquad
  r_\lambda=-\frac{\lambda^2+\omega_Q^2}{\alpha}, \label{eq: eigvecs_app}
\end{align}
so that $Ev_{\lambda}=(v_\lambda)_S=(1,\lambda)$, with
$c_\lambda=w_\lambda^\dagger \mathbf Dw_\lambda=\frac{\alpha^2D_0+D_2r_\lambda^2}{4|\lambda|^2(1+r_\lambda^2)^2}>0$. Then
\begin{align}
  \bar\Sigma_{L,S}&=\sum_\lambda c_\lambda\begin{pmatrix}1&\bar\lambda\\\lambda&|\lambda|^2\end{pmatrix}, \label{eq: Sigma_bar_S_app}
\end{align}
whose off-diagonal entries cancel, leaving
\begin{align}
  \bar\sigma_{11}&=\sum_\pm\frac{\alpha^2D_0+D_2r_\pm^2}{2\omega_\pm^2(1+r_\pm^2)^2},\qquad
  \bar\sigma_{22}=\sum_\pm\frac{\alpha^2D_0+D_2r_\pm^2}{2(1+r_\pm^2)^2},\qquad
  r_\pm=\frac{\omega_\pm^2-\omega_Q^2}{\alpha}. \label{eq: sigma_bar_entries_app}
\end{align}
Therefore,
\begin{align}
\det\Sigma_{L,S}&=t^2\bar\sigma_{11}\bar\sigma_{22}+O(t),\qquad
  \operatorname{Tr}\varrho_S^2\simeq \frac{1}{2t\sqrt{\bar\sigma_{11}\bar\sigma_{22}}}. \label{eq: detSigma_sub_app}
\end{align}

\emph{Critical}, $\alpha^2=\omega_Q^2\omega_c^2$, spectrum $\{0,\pm i\omega_+\}$:
$F$ is not diagonalisable and has a Jordan block for $\lambda=0$, so the spectral formula~\eqref{eq: Sigma_spectral} does not apply. Using generalised eigenvectors to form $\Pi_0=v_0w_0^{\mathsf T}+v_1w_1^{\mathsf T}$ and $N=F\Pi_0=v_0w_1^{\mathsf T}$, we have
$e^{Fs}=\Pi_0+sN+\sum_\pm e^{\pm i\omega_+s}\Pi_{\pm i\omega_+}\equiv\sum_{(\lambda,k)}s^k e^{\lambda s}\,C_{\lambda,k}$, with the indices $\lambda$ and $k$ referring to the projector and the associated power of $t$: the components are therefore $C_{0,0}=\Pi_0$, $C_{0,1}=N$ and $C_{\pm i\omega_+,0}=\Pi_\pm$. Inserting this into Eq.~\eqref{eq: Lyapunov_sol}, we obtain
\begin{align}
  \Sigma_L(t)&=\sum_{(\lambda,k),(\mu,l)}\Big[t^{k+l}e^{(\lambda+\mu)t}\,C_{\lambda,k}\Sigma_{L,0}C_{\mu,l}^{\mathsf T}
  +I_{k+l}(\lambda+\mu,t)\,C_{\lambda,k}\mathbf DC_{\mu,l}^{\mathsf T}\Big],
  \qquad I_n(m,t)=\int_0^ts^ne^{ms}ds, \label{eq: Sigma_crit_general}
\end{align}
where the integrals needed are
\begin{align}
  I_n(0,t)&=\frac{t^{n+1}}{n+1}, \quad 0\leq n\leq 2,\qquad
  I_0(m,t)=\frac{e^{mt}-1}{m},\qquad
  I_1(m,t)=\frac{t\,e^{mt}}{m}-\frac{e^{mt}-1}{m^2}\quad(m\neq0). \label{eq: I_integrals}
\end{align}
Let us now consider the zero-zero, the $i\omega_+$ and the mixed sectors one by one.

\emph{The zero-zero sector}:
With the right and left eigenvectors
\begin{align}
  v_0&=(1,0,r_0,0)^{\mathsf T},\quad
  v_1=(0,1,0,r_0)^{\mathsf T},\quad
  v_\pm=(1,\pm i\omega_+,r_+,\pm i\omega_+r_+)^{\mathsf T}, \label{eq: v_crit}\\
  w_0&=\frac{(1,0,r_0,0)^{\mathsf T}}{1+r_0^2},\quad
  w_1=\frac{(0,1,0,r_0)^{\mathsf T}}{1+r_0^2},\quad
  w_\pm=\frac{(\pm i\omega_+,1,\pm i\omega_+r_+,r_+)^{\mathsf T}}{\pm2i\omega_+(1+r_+^2)}, \label{eq: w_crit}
\end{align}
where $w_1$ is the left eigenvector, $w_1^{\mathsf T}F=0$, and $w_0^{\mathsf T}F=w_1^{\mathsf T}$, and
\begin{align}
  r_0&=-\frac{\omega_Q^2}{\alpha}=-\frac{\omega_Q}{\omega_c},\qquad
  r_+=\frac{\omega_+^2-\omega_Q^2}{\alpha}=\frac{\omega_c}{\omega_Q},\qquad r_0r_+=-1, \label{eq: r_crit}
\end{align}
we have, using $E=(I_2\;0)$, $Ev_0=e_1$, $Ev_1=e_2$ and $Ev_\pm=(1,\pm i\omega_+)^{\mathsf T}$.
Since $\mathbf D=\operatorname{diag}(0,\alpha^2D_0,0,D_2)$,
$w_0^{\mathsf T}\mathbf Dw_0=w_0^{\mathsf T}\mathbf Dw_1=w_0^{\mathsf T}\mathbf Dw_\pm=0$. Using
\begin{align}
E\Sigma_L^{00}E^{\mathsf T}&=E(\Pi_0+tN)\Sigma_{L,0}(\Pi_0+tN)^{\mathsf T}E^{\mathsf T}+E\big[t\,\Pi_0\mathbf D\Pi_0^{\mathsf T}+\tfrac{t^2}{2}(N\mathbf D\Pi_0^{\mathsf T}+\Pi_0\mathbf DN^{\mathsf T})+\tfrac{t^3}{3}N\mathbf DN^{\mathsf
  T}\big]E^{\mathsf T}, \label{eq: Sigma00_general}
\end{align}
we obtain
\begin{align}
  \Sigma_{L,S}^{(00)}(t)&=
  \frac{\omega_c^4}{(\omega_Q^2+\omega_c^2)^2}
  \begin{pmatrix}
  P_{0,11}+2t\,P_{0,12}+t^2P_{0,22} & P_{0,12}+t\,P_{0,22}\\[2pt]
  P_{0,12}+t\,P_{0,22} & P_{0,22}
  \end{pmatrix}
  +c\begin{pmatrix}t^3/3&t^2/2\\t^2/2&t\end{pmatrix}, \label{eq: Sigma00}
\end{align}
with
\begin{align}
  c&=w_1^{\mathsf T}\mathbf Dw_1=\frac{\alpha^2D_0+D_2r_0^2}{(1+r_0^2)^2}
  =\frac{\omega_Q^2\omega_c^2\,(\omega_c^4D_0+D_2)}{(\omega_Q^2+\omega_c^2)^2}. \label{eq: c_crit}
\end{align}

\emph{The $i\omega_+,i\omega_+$ sector}:
The initial-state part is
$\sum_{\pm,\pm'}e^{(\pm i\omega_+\pm'i\omega_+)t}E\Pi_\pm\Sigma_{L,0}\Pi_{\pm'}^{\mathsf T}E^{\mathsf T}$, and the noise
part is $\sum_{\pm,\pm'}I_0(\pm i\omega_+\pm'i\omega_+,t)\,E\Pi_\pm \mathbf D\Pi_{\pm'}^{\mathsf T}E^{\mathsf T}$. Only
the opposite-sign noise pairs are secular, $I_0(0,t)=t$. Thus
\begin{align}
  \Sigma_{L,S}^{(\pm\pm)}(t)&=t\,\bar\Sigma_{L,+,S}+B'(t),\qquad
  \bar\Sigma_{L,+,S}=\begin{pmatrix}s_+/\omega_+^2&0\\0&s_+\end{pmatrix},
  \qquad s_+=\frac{\omega_Q^2\omega_c^2(\omega_Q^4D_0+D_2)}{2(\omega_Q^2+\omega_c^2)^2}. \label{eq: Sigma_pp_crit}
\end{align}
The initial state enters only the bounded part.

\emph{The mixed sector}: The
initial-state part is $\sum_\pm e^{\pm i\omega_+t}E(\Pi_0+tN)\Sigma_{L,0}\Pi_\pm^{\mathsf T}E^{\mathsf T}$, and
the noise part is $\sum_\pm E\big[I_0(\pm i\omega_+,t)\Pi_0\mathbf D\Pi_\pm^{\mathsf T}+I_1(\pm i\omega_+,t)N\mathbf D\Pi_\pm^{\mathsf
  T}\big]E^{\mathsf T}$, each together with its transpose. Carrying out the sums over $\pm$ and the integrals gives, exactly,
\begin{align}
  \Sigma_{L,S}^{(0\pm)}(t)&=
  \frac{\omega_Q^2\omega_c^2}{(\omega_Q^2+\omega_c^2)^2}
  \left[
  \begin{pmatrix}1&t\\0&1\end{pmatrix}P_0
  \begin{pmatrix}\cos\omega_+t&-\omega_+\sin\omega_+t\\ \dfrac{\sin\omega_+t}{\omega_+}&\cos\omega_+t\end{pmatrix}
  +\begin{pmatrix}\cos\omega_+t&\dfrac{\sin\omega_+t}{\omega_+}\\ -\omega_+\sin\omega_+t&\cos\omega_+t\end{pmatrix}P_0
  \begin{pmatrix}1&0\\t&1\end{pmatrix}
  \right]
  \notag\\[4pt]
  &+\frac{\omega_Q^2\omega_c^2\,(\omega_Q^2\omega_c^2D_0-D_2)}{(\omega_Q^2+\omega_c^2)^2}
  \begin{pmatrix}
  \dfrac{2(\sin\omega_+t-\omega_+t\cos\omega_+t)}{\omega_+^3} & \dfrac{t\sin\omega_+t}{\omega_+}\\[10pt]
  \dfrac{t\sin\omega_+t}{\omega_+} & \dfrac{2\sin\omega_+t}{\omega_+}
  \end{pmatrix}. \label{eq: Sigma_0pm_crit}
\end{align}
Written out, the initial-state matrix has entries
\begin{align}
  (1,1)&:\ 2\Big[(P_{0,11}+tP_{0,12})\cos\omega_+t+\frac{P_{0,12}+tP_{0,22}}{\omega_+}\sin\omega_+t\Big], \label{eq: Sigma_0pm_11}\\
  (1,2)&:\ (2P_{0,12}+tP_{0,22})\cos\omega_+t+\Big(\frac{P_{0,22}}{\omega_+}-\omega_+(P_{0,11}+tP_{0,12})\Big)\sin\omega_+t, \label{eq: Sigma_0pm_12}\\
  (2,2)&:\ 2\big(P_{0,22}\cos\omega_+t-\omega_+P_{0,12}\sin\omega_+t\big), \label{eq: Sigma_0pm_22}
\end{align}
each multiplied by $\omega_Q^2\omega_c^2/(\omega_Q^2+\omega_c^2)^2$. Therefore,
\begin{align}
  \Sigma_{L,S}^{(0\pm)}&=\begin{pmatrix}O(t)&O(t)\\O(t)&O(1)\end{pmatrix}. \label{eq: Sigma_0pm_order}
\end{align}

Collecting terms, we have
\begin{align}
  \Sigma_{L,S}(t)&=\begin{pmatrix}c\,t^3/3+O(t^2)&c\,t^2/2+O(t)\\[2pt]c\,t^2/2+O(t)&(c+s_+)\,t+O(1)\end{pmatrix}, \label{eq: Sigma_S_crit}
\end{align}
so
\begin{align}
  \det\Sigma_{L,S}&=\Big[\frac{c\,t^3}{3}(c+s_+)t-\frac{c^2t^4}{4}\Big]+O(t^3)
  =c\Big(\frac c{12}+\frac{s_+}3\Big)t^4+O(t^3). \label{eq: detSigma_crit_derivation}
\end{align}
Explicitly,
\begin{align}
\det\Sigma_{L,S}&\simeq\frac{(\omega_Q^2\omega_c^2)^2}{(\omega_Q^2+\omega_c^2)^4}\,
  (\omega_c^4D_0+D_2)\,\frac{(\omega_c^4D_0+D_2)+2(\omega_Q^4D_0+D_2)}{12}\;t^4,
  \qquad
  \operatorname{Tr}\varrho_S^2=\frac{1}{2\sqrt{\det\Sigma_{L,S}}}\propto t^{-2}. \label{eq: detSigma_crit_app}
\end{align}
The leading coefficient is independent of the initial state.

At and above
$\alpha^2=\omega_Q^2\omega_c^2$, the dominant term compresses to rank one, along $(v_0)_S=(1,0)$ or
$(v_\kappa)_S=(1,\kappa)$, and $\det\Sigma_{L,S}$ is assembled from the orthogonal direction, giving the
$t^4$ and $t\,e^{2\kappa t}$ laws with the coefficients above.

\emph{Supercritical}, $\alpha^2>\omega_Q^2\omega_c^2$, spectrum $\{\pm i\omega_+,\pm\kappa\}$:
The eigenvectors are
\begin{align}
  v_\lambda&=(1,\lambda,r_\lambda,\lambda r_\lambda)^{\mathsf T},\qquad
  w_\lambda=\frac{(\lambda,1,\lambda r_\lambda,r_\lambda)^{\mathsf T}}{2\lambda(1+r_\lambda^2)},\qquad
  r_\lambda=-\frac{\lambda^2+\omega_Q^2}{\alpha}. \label{eq: eigvecs_super_app}
\end{align}
With $E=(I_2\;0)$, $Ev_\lambda=(1,\lambda)^{\mathsf T}$. One can again write $\Sigma_{L,S}$ sector by sector (i.e., the real, the mixed and the imaginary sectors).

\emph{The $\pm\kappa$ sector}:
\begin{align}
  \Sigma_{L,S}^{(\kappa\kappa)}(t)&=
  \frac{(\omega_c^2+\kappa^2)^2}{(\omega_+^2+\kappa^2)^2}
  \begin{pmatrix}\cosh\kappa t&\dfrac{\sinh\kappa t}{\kappa}\\ \kappa\sinh\kappa t&\cosh\kappa t\end{pmatrix}P_0
  \begin{pmatrix}\cosh\kappa t&\kappa\sinh\kappa t\\ \dfrac{\sinh\kappa t}{\kappa}&\cosh\kappa t\end{pmatrix}
  \notag\\[4pt]
  &+\frac{\alpha^2\big[D_0(\omega_c^2+\kappa^2)^2+D_2\big]}{(\omega_+^2+\kappa^2)^2}
  \begin{pmatrix}
  \dfrac{\sinh2\kappa t-2\kappa t}{4\kappa^3} & \dfrac{\sinh^2\kappa t}{2\kappa^2}\\[10pt]
  \dfrac{\sinh^2\kappa t}{2\kappa^2} & \dfrac{\sinh2\kappa t+2\kappa t}{4\kappa}
  \end{pmatrix}. \label{eq: Sigma_kk}
\end{align}

\emph{The mixed sector}:
\begin{align}
  \Sigma_{L,S}^{(\kappa+)}(t)&=
  \frac{\alpha^2}{(\omega_+^2+\kappa^2)^2}
  \left[
  \begin{pmatrix}\cosh\kappa t&\dfrac{\sinh\kappa t}{\kappa}\\ \kappa\sinh\kappa t&\cosh\kappa t\end{pmatrix}P_0
  \begin{pmatrix}\cos\omega_+t&-\omega_+\sin\omega_+t\\ \dfrac{\sin\omega_+t}{\omega_+}&\cos\omega_+t\end{pmatrix}
  +\begin{pmatrix}\cos\omega_+t&\dfrac{\sin\omega_+t}{\omega_+}\\ -\omega_+\sin\omega_+t&\cos\omega_+t\end{pmatrix}P_0
  \begin{pmatrix}\cosh\kappa t&\kappa\sinh\kappa t\\ \dfrac{\sinh\kappa t}{\kappa}&\cosh\kappa t\end{pmatrix}
  \right]
  \notag\\[4pt]
  &+\frac{\alpha^2(\alpha^2D_0-D_2)}{(\omega_+^2+\kappa^2)^2}
  \begin{pmatrix}
  \dfrac{2(\kappa\cosh\kappa t\sin\omega_+t-\omega_+\sinh\kappa t\cos\omega_+t)}{\kappa\omega_+(\kappa^2+\omega_+^2)} &
  \dfrac{\sinh\kappa t\,\sin\omega_+t}{\kappa\omega_+}\\[12pt]
  \dfrac{\sinh\kappa t\,\sin\omega_+t}{\kappa\omega_+} &
  \dfrac{2(\kappa\sinh\kappa t\cos\omega_+t+\omega_+\cosh\kappa t\sin\omega_+t)}{\kappa^2+\omega_+^2}
  \end{pmatrix}. \label{eq: Sigma_kp}
\end{align}
Every entry oscillates at $\omega_+$ with an amplitude growing like $e^{\kappa t}$, so this sector is
$O(e^{\kappa t})$.

\emph{The imaginary sector}:
\begin{align}
  \Sigma_{L,S}^{(++)}(t)&=
  \frac{(\omega_Q^2+\kappa^2)^2}{(\omega_+^2+\kappa^2)^2}
  \begin{pmatrix}\cos\omega_+t&\dfrac{\sin\omega_+t}{\omega_+}\\ -\omega_+\sin\omega_+t&\cos\omega_+t\end{pmatrix}P_0
  \begin{pmatrix}\cos\omega_+t&-\omega_+\sin\omega_+t\\ \dfrac{\sin\omega_+t}{\omega_+}&\cos\omega_+t\end{pmatrix}
  \notag\\[4pt]
  &+\frac{\alpha^2\big[D_0(\omega_Q^2+\kappa^2)^2+D_2\big]}{(\omega_+^2+\kappa^2)^2}
  \begin{pmatrix}
  \dfrac{2\omega_+t-\sin2\omega_+t}{4\omega_+^3} & \dfrac{\sin^2\omega_+t}{2\omega_+^2}\\[10pt]
  \dfrac{\sin^2\omega_+t}{2\omega_+^2} & \dfrac{2\omega_+t+\sin2\omega_+t}{4\omega_+}
  \end{pmatrix}. \label{eq: Sigma_pp_super}
\end{align}
This is a bounded initial-state term plus a
secular term growing like $t$.

Adding the three sectors,
\begin{align}
  \Sigma_{L,S}(t)&=A(t)\,P_0\,A(t)^{\mathsf T}+N(t), \label{eq: Sigma_S_super}
\end{align}
with
\begin{align}
  A(t)&=\frac{1}{\omega_+^2+\kappa^2}
  \begin{pmatrix}
  (\omega_c^2+\kappa^2)\cosh\kappa t+(\omega_Q^2+\kappa^2)\cos\omega_+t &
  (\omega_c^2+\kappa^2)\dfrac{\sinh\kappa t}{\kappa}+(\omega_Q^2+\kappa^2)\dfrac{\sin\omega_+t}{\omega_+}\\[10pt]
  (\omega_c^2+\kappa^2)\,\kappa\sinh\kappa t-(\omega_Q^2+\kappa^2)\,\omega_+\sin\omega_+t &
  (\omega_c^2+\kappa^2)\cosh\kappa t+(\omega_Q^2+\kappa^2)\cos\omega_+t
  \end{pmatrix}. \label{eq: A_super}
\end{align}
The noise matrix $N(t)$ has entries
\begin{align}
  N_{11}&=\frac{\alpha^2}{(\omega_+^2+\kappa^2)^2}\bigg\{
  \big[D_0(\omega_c^2+\kappa^2)^2+D_2\big]\frac{\sinh2\kappa t-2\kappa t}{4\kappa^3}
  +(\alpha^2D_0-D_2)\frac{2(\kappa\cosh\kappa t\sin\omega_+t-\omega_+\sinh\kappa t\cos\omega_+t)}{\kappa\omega_+(\omega_+^2+\kappa^2)}
  \notag\\
  &+\big[D_0(\omega_Q^2+\kappa^2)^2+D_2\big]\frac{2\omega_+t-\sin2\omega_+t}{4\omega_+^3}\bigg\}, \label{eq: N11}
  \\[4pt]
  N_{12}&=\frac{\alpha^2}{(\omega_+^2+\kappa^2)^2}\bigg\{
  \big[D_0(\omega_c^2+\kappa^2)^2+D_2\big]\frac{\sinh^2\kappa t}{2\kappa^2}
  +(\alpha^2D_0-D_2)\frac{\sinh\kappa t\,\sin\omega_+t}{\kappa\omega_+}
  +\big[D_0(\omega_Q^2+\kappa^2)^2+D_2\big]\frac{\sin^2\omega_+t}{2\omega_+^2}\bigg\}, \label{eq: N12}
  \\[4pt]
  N_{22}&=\frac{\alpha^2}{(\omega_+^2+\kappa^2)^2}\bigg\{
  \big[D_0(\omega_c^2+\kappa^2)^2+D_2\big]\frac{\sinh2\kappa t+2\kappa t}{4\kappa}
  +(\alpha^2D_0-D_2)\frac{2(\kappa\sinh\kappa t\cos\omega_+t+\omega_+\cosh\kappa t\sin\omega_+t)}{\omega_+^2+\kappa^2}
  \notag\\
  &+\big[D_0(\omega_Q^2+\kappa^2)^2+D_2\big]\frac{2\omega_+t+\sin2\omega_+t}{4\omega_+}\bigg\}. \label{eq: N22}
\end{align}

As $t\to\infty$, $\cosh\kappa t$ and $\sinh\kappa t$ each
become $\tfrac12e^{\kappa t}+O(e^{-\kappa t})$. Substituting into the three entries and sorting by
order gives
\begin{align}
  \Sigma_{L,S,11}&=\gamma\,e^{2\kappa t}+f_{11}(t)\,e^{\kappa t}
  +\frac{\alpha^2}{(\omega_+^2+\kappa^2)^2}\Big[\frac{D_0(\omega_Q^2+\kappa^2)^2+D_2}{2\omega_+^2}-\frac{D_0(\omega_c^2+\kappa^2)^2+D_2}{2\kappa^2}\Big]t+O(
  1), \label{eq: SigmaS11_asym}
  \\
  \Sigma_{L,S,12}&=\kappa\,\gamma\,e^{2\kappa t}+f_{12}(t)\,e^{\kappa t}+O(1), \label{eq: SigmaS12_asym}
  \\
  \Sigma_{L,S,22}&=\kappa^2\gamma\,e^{2\kappa t}+f_{22}(t)\,e^{\kappa t}
  +\frac{\alpha^2}{(\omega_+^2+\kappa^2)^2}\Big[\frac{D_0(\omega_Q^2+\kappa^2)^2+D_2}{2}+\frac{D_0(\omega_c^2+\kappa^2)^2+D_2}{2}\Big]t+O(1), \label{eq: SigmaS22_asym}
\end{align}
with
\begin{align}
  \gamma&=\frac{1}{(\omega_+^2+\kappa^2)^2}\left[
  \frac{(\omega_c^2+\kappa^2)^2\big(\kappa^2P_{0,11}+2\kappa P_{0,12}+P_{0,22}\big)}{4\kappa^2}
  +\frac{\alpha^2\big[D_0(\omega_c^2+\kappa^2)^2+D_2\big]}{8\kappa^3}\right], \label{eq: gamma_app}
\end{align}
\begin{align}
  \tilde K&=\frac{\alpha^2(\alpha^2D_0-D_2)}{(\omega_+^2+\kappa^2)^2}, \label{eq: Ktilde}
\end{align}
and
\begin{align}
  f_{11}(t)&=\frac{\alpha^2}{(\omega_+^2+\kappa^2)^2}\bigg[
  \frac{(\kappa P_{0,11}+P_{0,12})\cos\omega_+t}{\kappa}+\frac{(\kappa P_{0,12}+P_{0,22})\sin\omega_+t}{\kappa\,\omega_+}\bigg]
  +\tilde K\,\frac{\kappa\sin\omega_+t-\omega_+\cos\omega_+t}{\kappa\,\omega_+(\omega_+^2+\kappa^2)}, \label{eq: f11}
  \\[4pt]
  f_{12}(t)&=\frac{\alpha^2}{2(\omega_+^2+\kappa^2)^2}\bigg[
  \Big(\kappa P_{0,11}+2P_{0,12}+\frac{P_{0,22}}{\kappa}\Big)\cos\omega_+t
  +\Big(\frac{\kappa P_{0,12}+P_{0,22}}{\omega_+}-\frac{\omega_+(\kappa P_{0,11}+P_{0,12})}{\kappa}\Big)\sin\omega_+t\bigg]\notag\\
  &+\tilde K\,\frac{\sin\omega_+t}{2\kappa\,\omega_+}, \label{eq: f12}
  \\[4pt]
  f_{22}(t)&=\frac{\alpha^2}{(\omega_+^2+\kappa^2)^2}\Big[
  (\kappa P_{0,12}+P_{0,22})\cos\omega_+t-\omega_+(\kappa P_{0,11}+P_{0,12})\sin\omega_+t\Big]
  +\tilde K\,\frac{\kappa\cos\omega_+t+\omega_+\sin\omega_+t}{\omega_+^2+\kappa^2}. \label{eq: f22}
\end{align}

These satisfy
\begin{align}
  \kappa^2f_{11}^{\rm noise}-2\kappa f_{12}^{\rm noise}+f_{22}^{\rm noise}
  &\propto\kappa^2\sin\omega_+t-\kappa\omega_+\cos\omega_+t-(\omega_+^2+\kappa^2)\sin\omega_+t+\kappa\omega_+\cos\omega_+t+\omega_+^2\sin\omega_+t=0 . \label{eq: f_noise_identity}
\end{align}
We also have
\begin{align}
  \kappa^2f_{11}-2\kappa f_{12}+f_{22}&=0 . \label{eq: f_identity}
\end{align}

\emph{The determinant}: Multiplying out $\Sigma_{L,S,11}\Sigma_{L,S,22}-\Sigma_{L,S,12}^2$ order by order,
\begin{align}
  e^{4\kappa t}:&\quad \gamma\cdot\kappa^2\gamma-(\kappa\gamma)^2=0, \label{eq: det_order_4}\\
  e^{3\kappa t}:&\quad \gamma\big(\kappa^2f_{11}-2\kappa f_{12}+f_{22}\big)=0, \label{eq: det_order_3}\\
  t\,e^{2\kappa t}:&\quad \gamma\Big(\kappa^2[\text{$t$-coefficient of }\Sigma_{L,S,11}]+[\text{$t$-coefficient of }\Sigma_{L,S,22}]\Big)\notag\\
  &\quad=\gamma\,\frac{\alpha^2}{(\omega_+^2+\kappa^2)^2}\Big[\big(D_0(\omega_Q^2+\kappa^2)^2+D_2\big)\Big(\frac{\kappa^2}{2\omega_+^2}+\frac12\Big)
  +\big(D_0(\omega_c^2+\kappa^2)^2+D_2\big)\Big(-\frac{\kappa^2}{2\kappa^2}+\frac12\Big)\Big]\notag\\
  &\quad=\gamma\,\frac{\alpha^2\big[D_0(\omega_Q^2+\kappa^2)^2+D_2\big]}{2\omega_+^2(\omega_+^2+\kappa^2)}, \label{eq: det_order_2}
\end{align}
and everything else is $O(e^{2\kappa t})$. Therefore
\begin{align}
\det\Sigma_{L,S}&=\gamma\,s_\perp\,t\,e^{2\kappa t}\big(1+O(t^{-1})\big),\qquad
  s_\perp=\frac{\alpha^2\big[D_0(\omega_Q^2+\kappa^2)^2+D_2\big]}{2\omega_+^2(\omega_+^2+\kappa^2)},
\qquad\operatorname{Tr}\varrho_S^2\simeq\frac{1}{2\sqrt{\gamma s_\perp}}\,\frac{e^{-\kappa t}}{\sqrt t}. \label{eq: detSigma_super_app}
\end{align}

\end{document}